\documentclass[usenatbib]{mn2e}
\usepackage{multirow}
\usepackage{rotating}
\usepackage{colortbl}
\usepackage{color}
\usepackage[fleqn]{amsmath}
\usepackage{amssymb}
\usepackage{amsfonts}
\usepackage{verbatim}
\usepackage{scalefnt}
\usepackage[percent]{overpic}

\newcommand{\apjl}{ApJ}
\newcommand{\apjs}{ApJS}
\newcommand{\ao}{Appl.~Opt.}
\newcommand{\aap}{A\&A}

\newcommand{\beq}{\begin{equation}}
\newcommand{\eeq}{\end{equation}}

\newcommand{\msun}{\,{\rm M_\odot}}
\definecolor{grey}{rgb}{0.5,0.6,0.7}
\def \simlt { \lower .75ex \hbox{$\sim$} \llap{\raise .27ex \hbox{$<$}} }
\definecolor{purple}{rgb}{0.65,0.15,0.9}
\definecolor{darkorange}{rgb}{0.8,0.3,0}
\definecolor{olive}{rgb}{0.4,0.6,0.25}
\definecolor{darkgreen}{rgb}{0,0.7,0}
\definecolor{darkred}{rgb}{0.5,0,0}

\title[Blossoms from black hole seeds]{Blossoms from black hole seeds: properties and early growth regulated by supernova feedback
}
\author[Habouzit et al.]{M\'{e}lanie Habouzit$^{1}$\thanks{E-mail: habouzit@iap.fr},
 Marta Volonteri$^{1}$,
 Yohan Dubois$^{1}$\\
 $^1$Institut d'Astrophysique de Paris, Sorbonne Universit\'{e}s, UPMC Univ Paris 6 et CNRS, UMR 7095, 98 bis bd Arago, 75014 Paris, France\\}

\begin{document}
\maketitle

\begin{abstract}
Massive black holes (BHs) inhabit local galaxies, including the Milky Way and some dwarf galaxies. BH formation, occurring at early cosmic times, must account for the properties of BHs in today's galaxies, notably why some galaxies host a BH, and others do not.
We investigate the formation, distribution and growth of BH `seeds' by using the adaptive mesh refinement code {\sc Ramses}.
We develop an implementation of BH formation in dense, low-metallicity environments, as advocated by models invoking the collapse of the first generation of stars, or of dense nuclear star clusters.  The seed masses are computed one-by-one on-the-fly, based on the star formation rate and the stellar initial mass function.  This self-consistent method to seed BHs allows us to study the distribution of BHs in a cosmological context and their evolution over cosmic time. We find that all high-mass galaxies tend to host a BH, whereas low-mass counterparts have a lower probability of hosting a BH. After the end of the epoch of BH formation, this probability is modulated by the growth of the galaxy.   
The simulated BHs connect to low-redshift observational samples, and span a similar range in accretion properties as Lyman-Break Analogs. The growth of BHs in low-mass galaxies is stunted by strong supernova feedback.  The properties of BHs in dwarf galaxies thus remain a testbed for BH formation. Simulations with strong supernova feedback, which is able to quench BH accretion in shallow potential wells,  produce galaxies and BHs in better agreement with observational constraints.

\end{abstract}

\begin{keywords}
{black hole physics, galaxies:high redshift, galaxies:formation, galaxies:evolution, methods:numerical}

\end{keywords}

\section{Introduction}
\label{sec:intro}

Massive black holes (BH) of millions of solar masses and above reside in the centre of most local galaxies, including the Milky Way. BHs are also observed in low-mass galaxies \citep{2012NatCo...3E1304G,2013ApJ...775..116R,2015ApJ...799...98M}. However we also observe galaxies without any signatures of BHs \citep[e.g., M33, NGC~205,][]{Gebhardt2001,Valluri2005}. BHs  power active galactic nuclei (AGN) and quasars, therefore driving feedback onto their host galaxies, believed to be key in shaping the massive end of the galaxy stellar mass function \citep{Croton2006,Silk2013}. Several empirical relations have been established between BH masses and their host galaxy properties   \citep[total stellar mass, stellar mass in the bulge, velocity dispersion of the galaxy, see][and references therein]{2013ARA&A..51..511K}. BHs appear to be a key element in galaxy evolution, but the question of their formation is still an open one. What fraction of galaxies hosts BHs? What is the minimum galaxy mass below which BHs become rare? How do BHs form?

Currently, three main scenarios are popular to explain theoretically the formation of such massive seeds in the early Universe \citep[see][for a seminal discussion on BH formation]{Rees1978}.
In the ``PopIII star remnants'' scenario, BHs are predicted to form in mini-haloes ($\rm{M_{h}}\approx \, 10^{5}\, \rm{M_{\odot}}$) with gas below a critical metallicity  \citep[$Z<10^{-3.5} \, \rm{Z_\odot}$,][]{Bromm01,Schneider2002} at redshift $z=30-20$ from the remnants of the first generation of stars  \citep[PopIII,][]{MadauRees2001,VMH}. Observational evidence on the initial mass function (IMF) of PopIII stars are lacking, but theoretical studies suggest that they could have masses in the range $10-1000 \,\rm{M_{\odot}}$ \citep{2011ARA&A..49..373B, Hirano2014}. A massive star $\rm{M_{\star}} \gtrsim \, 260 \rm{M_{\odot}}$ can lead to the formation of a BH seed of $\approx 100\, \rm{M_{\odot}}$ \citep{2001ApJ...550..372F}, retaining half the mass of the star.

Compact nuclear clusters often inhabit the centre of galaxies. Such a cluster, in the relatively metal-poor environments of high redshift galaxies, could have collapsed and formed a very massive star, up to $\sim 1000\, \rm{M_{\odot}}$, by stellar collisions. In metal-poor conditions mass losses through stellar winds are limited and massive remnants possible. \cite{Yungelson2008} studied the stellar evolution of solar composition stars in the mass range $60-1000 \, \rm{M_{\odot}}$. They found that they shed most of their mass via winds and are expected to end their lives as BHs less massive than $\sim 150\, \rm{M_{\odot}} $. At low metallicity, instead, mass loss due to winds is much reduced, thus increasing the mass of the final remnant \citep{Heger2003,2008NewAR..52..419V,2009MNRAS.395L..71M,2010ApJ...715L.138B,2013MNRAS.429.2298M,2015MNRAS.451.4086S}. A BH seed of $10^3\, \rm{M_{\odot}}$ can be formed in this ``nuclear cluster" scenario \citep{Omukai2008,Devecchi2009,Regan09,Katz2015}. In this paper we implement a model for BH seed formation which mimics the PopIII star remnants scenario and the nuclear cluster scenario.

Another scenario, which we have discussed in companion papers \citep{2016arXiv160100557H,2016MNRAS.456.1901H}, leads to the formation of $10^{4}-10^{6} \rm{M_{\odot}}$ BH seeds. The ``direct collapse''  of a single supermassive star into a massive BH requires very specific dynamical or thermodynamical conditions  \citep{Loeb1994,2003ApJ...596...34B,2004MNRAS.354..292K,spaans2006,Begelman2006,Lodato2006,Dijkstra2008,2008ApJ...682..745W,Regan09,2013MNRAS.433.1607L}, making it less common. To form only one supermassive star out of a gas cloud, the Jeans mass must remain large to avoid fragmentation. The Jeans mass increases with the gas temperature, so this model requires that all  efficient coolants (metals and molecular hydrogen) are absent. The only remaining coolant is atomic hydrogen, which cannot cool gas below $\sim 8000$~K, thus leading to a very large Jeans mass of $10^{5}\, \rm{M_{\odot}}$. In the very centre of a metal-free halo where molecular hydrogen formation is suppressed for at least 10~Myr \citep{2014arXiv1406.7020V,2015MNRAS.452.1026L}, and in the presence of large inflow rates ($>0.1\, \rm{M_{\odot}}/yr $), a supermassive star can form, and from it a BH, retaining up to $90 \%$ of the stellar mass. This models predicts massive, but rare seeds, which can possibly explain the quasar population at $z>6$ \citep{Fan2006,jiange09,Mortlock2011}, but has more difficulties with accounting for the presence of BHs in almost all galaxies today \citep{2016arXiv160100557H}. 

So far, studies have often focussed on addressing the high-mass end of the BH mass distribution, in order to explain the quasar population at $z=6$ and the powerful AGN at lower redshift which are expected to influence star formation in massive galaxies. In most cosmological simulations, BHs with mass $\sim 10^4-10^5\, \rm{M_{\odot}}$ are seeded in haloes above a fixed mass threshold, typically $\rm{M_{h}\sim 10^{10}\, \rm{M_{\odot}}}$ \citep[e.g.][]{Sijacki2009,2012ApJ...745L..29D,Hirschmann2012,2015MNRAS.452..575S}.  \citet{Bellovary2011}, \citet{2012MNRAS.420.2662D}, \citet{2014MNRAS.442.2751T}, \citet{,2014MNRAS.440.1590D} and \cite{2016arXiv160201941V} instead seed BHs in high gas density peaks, in some cases with a metallicity criterion, with fixed BH mass in the range $10 \leq \rm{M_{\odot}} \leq10^{5} \, \rm{M_{\odot}}$. Most of these simulations have been focused on studying the growth of BHs and AGN feedback in massive galaxies, rather than their formation, and have successfully reproduced the AGN luminosity function, which is dominated by BH with masses $\sim 10^8 \msun$. For such massive BHs the birth properties have been forgotten, and the details of BH formation are not crucial.

The imprint of BH formation is indeed not to be found in massive galaxies, where the central BH must have grown by several orders of magnitude.  Dwarf galaxies, instead, where neither the galaxy nor the BH can have grown much over cosmic time, provide us a promising laboratory where the mass of the central BH is expected to not differ much from its initial mass \citep{volonteri2008,2010MNRAS.408.1139V,2013ApJ...775..116R}. 
For example, a recent zoom cosmological hydrodynamical simulation from~\cite{2015MNRAS.452.1502D} has shown that a strong stellar feedback can suppress the growth of the BH until the galaxy has acquired enough mass at around $\sim10^9\, \rm M_\odot$.
Below this stellar mass, SN-driven winds are fast enough to overcome the escape velocity of the gravitational potential of the galaxy, and cold gas is routinely removed from the central parts of the galaxy.
Low-mass galaxies are also key to distinguish between formation scenarios through a different diagnostic, the occupation fraction, i.e. the probability that a galaxy of a given mass hosts a BH \citep{volonteri2008,2010MNRAS.408.1139V,2012NatCo...3E1304G}. The ``direct collapse" scenario, requiring very strict conditions, would leave many galaxies bereft of a BH, while less exacting models predict that a larger fraction of galaxies are eligible to host a BH. In principle, the mass and the occupation fraction of BHs in low-mass galaxies can therefore constrain BH formation. Of course, one should keep in mind that different models are not mutually exclusive in the Universe \citep[e.g.,][]{Volonteri2010,Devecchi2012,Lup2014}.

In this paper, we develop a new method to seed cosmological simulations with BHs. Our approach is based on the local gas and stellar properties, and captures the properties of both ``PopIII star remnants" and ``nuclear cluster" models. To test BH formation against observations, we compare our sample of BHs with a low-redshift sample of local galaxies (including broad-line AGN, galaxies with dynamical BH mass measurement, and several dwarf galaxies), and with Lyman-Break Analogs (LBAs). LBAs have similar properties to the more distant LBGs, but they are local systems that can be studied in much greater detail. 
We use the code {\sc ramses} \citep{Teyssier02}, which is a grid-based hydrodynamical solver with adaptive mesh refinement, that we describe in Section 2. Our new implementation of BH formation is detailed in Section 3. Results are presented in Sections 4, 5 and 6, and comparison with observations are detailed in Section 7

\section{Simulation parameters}
\subsection{Initial conditions}
\label{subsec:ramses}
We have performed three simulations, called SuperChunky, which only differ by the prescription of supernova (SN) feedback. These runs are a higher resolution version of Chunky, the simulation used by \citet{2016arXiv160100557H} to study  the number density of ``direct collapse" BHs, but they have a better dark matter resolution, thus allowing us to resolve smaller halos.
We use a $\Lambda$ cold dark matter cosmology, with total matter density $\Omega_{\rm m}=0.276$, dark matter energy density $\Omega_{\rm \Lambda}=0.724$, amplitude of the matter power spectrum $\sigma_{\rm 8}=0.811$, spectral index $n_{\rm s}=0.961$, baryon density $\Omega_{\rm b}=0.045$ and Hubble constant $\rm{H_{\rm 0}}= 70.3\, \rm{km\,  s^{-1} \, Mpc^{-1}}$, compatible with WMAP-7 \citep{2011ApJS..192...18K}.
Simulations are performed in a periodic box of side $10$ comoving $ \rm{Mpc}\, (cMpc)$ with $256^{3}$ dark matter particles,
corresponding to a mass resolution of $M_{\rm{DM,res}}=1.65 \times10^{6}\, \rm{M_{\odot}}$. Simulations are run from redshift $z=100$ to $z=2$. We use nested grid initial conditions built with the code {\sc music} \citep{2013ascl.soft11011H}. The simulations are run using the adaptive mesh refinement hydrodynamical cosmological code {\sc ramses} \citep[][version 2013]{Teyssier02}. 
Particles are projected on the grid with a cloud-in-cell interpolation and the Poisson equation is solved with an adaptive particle-mesh solver.
The Euler equations are solved with a MUSCL scheme using an approximate Harten-Lax-Van Leer Riemann solver, with a Min-Mod total variation diminishing scheme to linearly interpolate the cell-centered values to their edge locations.
Cells are refined (unrefined) based on a quasi-Lagrangian criterion: with more (less) than 8 DM particles in a cell, or with a total baryonic mass higher (smaller) than 8 times the DM mass resolution. To keep the refinement quasi homogeneous in physical units throughout cosmic time a new refinement level is allowed only when the expansion factor is doubled, namely for $a_{\rm exp}=0.1,0.2,0.4$ and so on.
The initial mesh is refined with 9 levels of refinement, 
leading to a spatial resolution of $\Delta x= 76\, \rm{pc}$. 

\subsection{Physics of the simulations}
Our simulations include sub-grid physics for cooling, star formation, SN feedback, AGN feedback.
Cooling is modeled with  the cooling curves of \cite{1993ApJS...88..253S}, the gas cools through H, He, and metals. The metallicity of the gas is modeled as a passive variable, which makes it easily trackable over the gas flow through redshift evolution. An initial zero metallicity is assumed for all the simulations in this work. Physical processes, such as SN explosions and star formation, modify and redistribute the metallicity over neighboring cells. 
To mimic reionization, heating from an uniform UV background is added (following \citealp{1996ApJ...461...20H}), taking place after $z=8.5$.
Star formation occurs in dense ($\rho>\rho_{0}$, with $\rho$ the density of the gas, $\rho_{0}$ the gas density threshold, see equation~\ref{eq:SF}) and cold ($T<T_{0}$, see equation~\ref{eq:TSF}) gas. The gas density threshold is set to  $\rho_{0}=1\, \rm{H\, cm}^{-3}$ for the three SuperChunky simulations. Star formation is modeled with a Kennicutt-Schmidt law:
\begin{equation}
\frac{d\rho_{\star}}{dt}=\epsilon_{\star}\frac{\rho}{t_{\rm{ff}}}
\label{eq:SF}
\end{equation}
with $\dot{\rho_{\star}}$ the star formation rate density, $\epsilon_{\star}=0.02$ the star formation efficiency (constant with redshift), and $t_{\rm{ff}}$ the free-fall time of the gas. Stars are created with a Poisson random process calculating the probability to form N stars with a mass resolution of  $\rm{m}_{\rm{res,\star}}= 7.7 \times 10^{3}\, \rm{M_{\odot}}$~\citep{rasera&teyssier06}.

The gas follows an adiabatic equation-of-state (EoS) for monoatomic gas with adiabatic index  $\gamma=5/3$, except at high gas densities $\rho>\rho_0$, where we use a polytropic EoS to increase the gas pressure in dense gas in order to limit excessive gas fragmentation by mimicking heating of the interstellar medium from stars~\citep{springel&hernquist03}:
\begin{equation}
T=T_{0}\left(\frac{\rho}{\rho_{0}}\right)^{\kappa-1}
\label{eq:TSF}
\end{equation}
with $T$ the gas temperature, $T_{0}$ the temperature threshold, $\rho_{0}$ the density threshold, and $\kappa$ the polytropic index of the gas. 
We use $\rm \kappa = 1.6 $ for the polytropic index, and $\rm T_{0}=10^{3}\, K$. 

\subsection{SN feedback}
We model type II SNe assuming a Chabrier 
initial mass function  \citep{2003PASP..115..763C}, where $\eta_{\rm{SN}} = 20\%$ of the mass fraction of stars end up their life in type II SNe, and release $e_{\rm SN}=10^{50} \, \rm erg\, M_\odot^{-1}$, and return metals with a yield of 0.1.

We employ three different SN feedback models implemented in {\sc ramses}: (i) a thermal feedback (simulation `T'), (ii) a kinetic feedback  (simulation `K') and (iii) a delayed cooling model (simulation `D'). 
We use a weak ``thermal" SN feedback which releases only internal energy in the neighboring cells \citep{2008A&A...477...79D}. 
The kinetic SN feedback \citep{2008A&A...477...79D} is modeled to reproduce a Sedov blast wave, where energy, mass and momentum are deposited in the neighboring gas cells.
The third model is called delayed cooling \citep{Stinson2006,Teyssier13}.
After a SN explosion, the coupling of the energy to the gas is not trivial to model, because the energy released by the explosion can be stored by non-thermal processes, such as unresolved turbulence, magnetic fields, cosmic rays, which are not captured in our simulations.  
These processes will dissipate their energy on potentially longer timescales, defined as the dissipative time $t_{\rm diss}$. In order to mimic the energetic and pressure enhancement by the non-thermal processes, in the delayed cooling implementation gas cooling is prevented in gas cells where the non-thermal energy component (or non-thermal velocity dispersion~$\sigma_{\rm{NT}}$) is larger than some user-defined threshold~\citep{Teyssier13}. We adopt the implementation of \citet{2015MNRAS.452.1502D}, where the two parameters $t_{\rm diss}$ and $\sigma_{\rm{NT}}$ are resolution-dependent. This allows us to adopt the values of $t_{\rm diss}$ and $\sigma_{\rm NT}$ that are required for the blast wave to propagate over a Jeans length (i.e. 4 high-resolution cells).
Moreover, to produce a more bursty SN feedback we explode only one stellar particle out of 10 stellar particles, but with ten times more SN specific energy.
For the present simulation SuperChunky, we use the parameters: $\sigma_{\rm NT}=65
\, \rm{km \, s^{-1}}$, $t_{\rm diss}=4.6 \, \rm{Myr}$ (see the Appendix A of \citealp{2015MNRAS.452.1502D} for a complete description of these two parameters). 

\subsection{Halo and galaxy finder codes}
We construct catalogues of haloes using the AdaptaHOP halo finder~\citep{Aubert+04}, which uses an SPH-like kernel to compute densities at the location of each particle and partitions the ensemble of particles into sub-haloes based on saddle points in the density field. Haloes contain at least 100 dark matter particles.  Galaxies are identified in the same way, and contain at least 100 stellar particles.

\section{Seeding cosmological simulations with BH seeds}
In this section, we describe our implementation to seed BHs in large-scale simulations, with an approach that is inspired by and mimics the PopIII star remnants and nuclear stellar cluster scenarios.  Regions where  BHs form are not identified on haloes properties, but on local environment properties \citep[see also][]{Bellovary2011}. BH masses are computed one by one, according to the gas and stellar properties of these regions.

\subsection{Selecting BH formation regions}
We modify  the clump finder routine in {\sc ramses} \citep[see][]{2014MNRAS.445.4015B}, which identifies regions denser than a given threshold. We use $\rho_{0}=\rho_{\star}$, i.e. the threshold for BH formation is the same as for star formation (as the models of BH formation we want to model are based on stars, rather than gas collapse). Thus, the formation of BHs happens in the same dense regions as those of star formation.
Two clumps are merged if they share a saddle point, which has a density higher than the density threshold.
We then verify several physical criteria: overdensities must be contracting along all axes, must be bound, and no pre-existing BH should exist within the overdensity. We also add a criterion on the metallicity of the gas in the clump, which is crucial to determine the formation rate of BHs in this type of models and its eventual dwindling \citep{Bellovary2011}. The metallicity of the clump is the mass-weighted metallicity of gas cells belonging to the clump. At this stage, collapsing regions with $Z<10^{-3.5} \, \rm{Z_\odot}$ are flagged as possible BH formation sites.

\citet{2014MNRAS.442.2751T} follows a similar approach, where cosmological simulations are seeded with PopIII remnant BHs if a gas particle density exceeds a given density threshold, and is metal-free ($Z=0$). However, their initial seed BH mass is fixed, whereas our model compute individually each seed BH mass.

\subsection{Computing BH initial masses}
Once metal-free collapsing regions are selected in our simulation box, we compute the theoretical mass in low-metallicity stars which can be formed in each clump using the Kennicutt-Schmidt law. We then calculate the probability of forming massive stars, adopting an IMF for the PopIII stars. Since we are focusing on low-metallicity stars ($Z<10^{-3.5}\, \rm{Z_\odot}$), we have considered a  logarithmically flat IMF, as suggested by investigations of the formation of PopIII stars \citep{Hirano2014}. The adopted minimum and maximum stellar mass are 1 and 500 $\, \rm{M_{\odot}}$ respectively. This IMF enters only in the implementation of BH formation and not in the SN feedback implementation.  
Defining  $\xi$ as the IMF per unit of stellar mass, $ \xi \equiv \rm{m} \frac{\mathrm dN}{\mathrm d\rm{m}}$,  the total mass in stars with masses between $\rm{m_{1}}$ and $\rm{m_{2}}$ is:
\begin{eqnarray}
\rm{M_*} = \int_{m_{1}}^{m_{2}}  m \, \Phi (m) \, \mathrm dm = \int_{m_{1}}^{m_{2}}  \xi (m) \, \mathrm dm\, .  
\end{eqnarray}

Specifically when we think of the PopIII remnants scenario, BH seeds are expected to form from stars in two mass ranges (Heger \& Woosley 2002): $25<\rm{m}<140\, \rm{M_{\odot}}$ and $260< \rm m<500\, \rm M_{\odot}$. The low-mass range is unlikely to form BHs eligible to become central BHs, as they are not sufficiently massive to remain in the galaxy centre \citep{Volonteri2010AARV}. The high-mass range is more favorable. If the stellar mass in clump is too small, however, the probability of forming a star with $\rm{m}>260 \, \rm{M_{\odot}}$ is smaller than unity. We stochastically sample the IMF and find that the probability of forming a sufficiently massive star becomes close to unity when the stellar mass in the clump is $\sim 10^3\, \rm{M_{\odot}}$. In our simulations gas clumps are always more massive than this value, therefore we can assume that the probability of BH formation is unity in a given clump. 

We define a parameter $f_{\rm{BH}}$ to describe the fraction of stellar mass which goes into the BH. We integrate the IMF (in mass) to compute the stellar mass fraction of stars within the range $260-500\, \rm M_{\odot}$, and find $f_{\rm BH}=0.48$.

To this, we add an efficiency $\epsilon_{\rm{BH}}$ that accounts for the ratio of the mass of the BH to its parent star, and we conservatively assume that a BH retains 50\% of the stellar mass, and we release metals in the surroundings accordingly.  The initial mass of the sink particle is finally expressed by $\rm{M}_{\rm{BH}}=f_{\rm{BH}} \times \epsilon_{\rm{BH}} \times \rm{M_{\star}} = 0.48 \times 0.50 \times \rm{M_{\star}} =0.24 \times  \rm{M_{\star}}$. 

For the nuclear cluster scenario the metallicity range is the key parameter allowing for formation of a BH with mass $\sim 10^3 \msun$ (see Devecchi \& Volonteri 2009), once a dense cluster of stars forms in the clump. As shown in section~4.1 below with the method described above we form already BHs with mass  $\sim 10^3 \msun$, in line with the expectations for this model. We therefore do not differentiate explicitly between the two models, such approach would require even higher resolution simulations that resolve clumps with mass $\ll 10^3 \msun$. Moreover, because we do not differentiate between the PopIII remnant and stellar cluster models, and that our model is an intermediate model between these two, we assume that all the BHs that could form merge together.

\subsection{BH growth and AGN feedback}
The accretion on the BH is described by the minimum between a Bondi-Hoyle-Lyttleton accretion rate and the Eddington accretion rate, with:
\begin{eqnarray}
\rm{\dot{{M}}_{BH}}=4\pi \alpha G^{2} \rm{{\bar \rho M_{BH}^{2}} \over \left( \bar c_s^{2}+\bar v^{2}\right)^{3/2}}
\end{eqnarray}
where $\alpha$ is a boost factor \citep{Booth2009} equals to $(\rho/\rho_0)^2$ when $\rho>\rho_0$ and 1 otherwise, G is the gravitational constant, $\rm{M_{BH}}$ the mass of the BH, $\bar \rho$ the average density of the medium, $\rm \bar c_s$ the average of the sound speed, $\rm \bar v$ the average velocity of the gas relative to the BH.
The accretion is conservatively limited to the Eddington accretion rate:
\begin{eqnarray}
\rm{\dot{{M}}_{Edd}}={4\pi G \rm{m_{p} M_{BH}} \over \epsilon_{r} \sigma_{T} c},
\end{eqnarray}
where $\rm m_{p}$ the proton mass, $\rm \epsilon_{r}=0.1$ the radiative efficiency, $\rm \sigma_{T}$ the Thomson cross-section, and $\rm c$ the speed of light.

AGN feedback is modeled with an isotropic injection of thermal energy into the surrounding gas, within a sphere of 4 cells ($4 \times \Delta x$) around sink particles \citep{2011MNRAS.414..195T,duboisetal12}. We store the rest-mass energy of the accreted mass into the BH, and release it when the energy is high enough to raise the temperature of the gas around the BH to at least $10^{7} \rm{K}$. The energy is released as thermal energy, with an efficiency of $\epsilon_{f}=0.15$ (calibrated to reproduce the observational $\rm{M_{BH}-M_{\star}}$ and $\rm{M_{BH}-\sigma_{\star}}$ relations), within a spherical bubble of 4 cells (4 $\times$ $\Delta x$) centred around the sink particle.

\begin{figure}
   \includegraphics[scale=0.4]{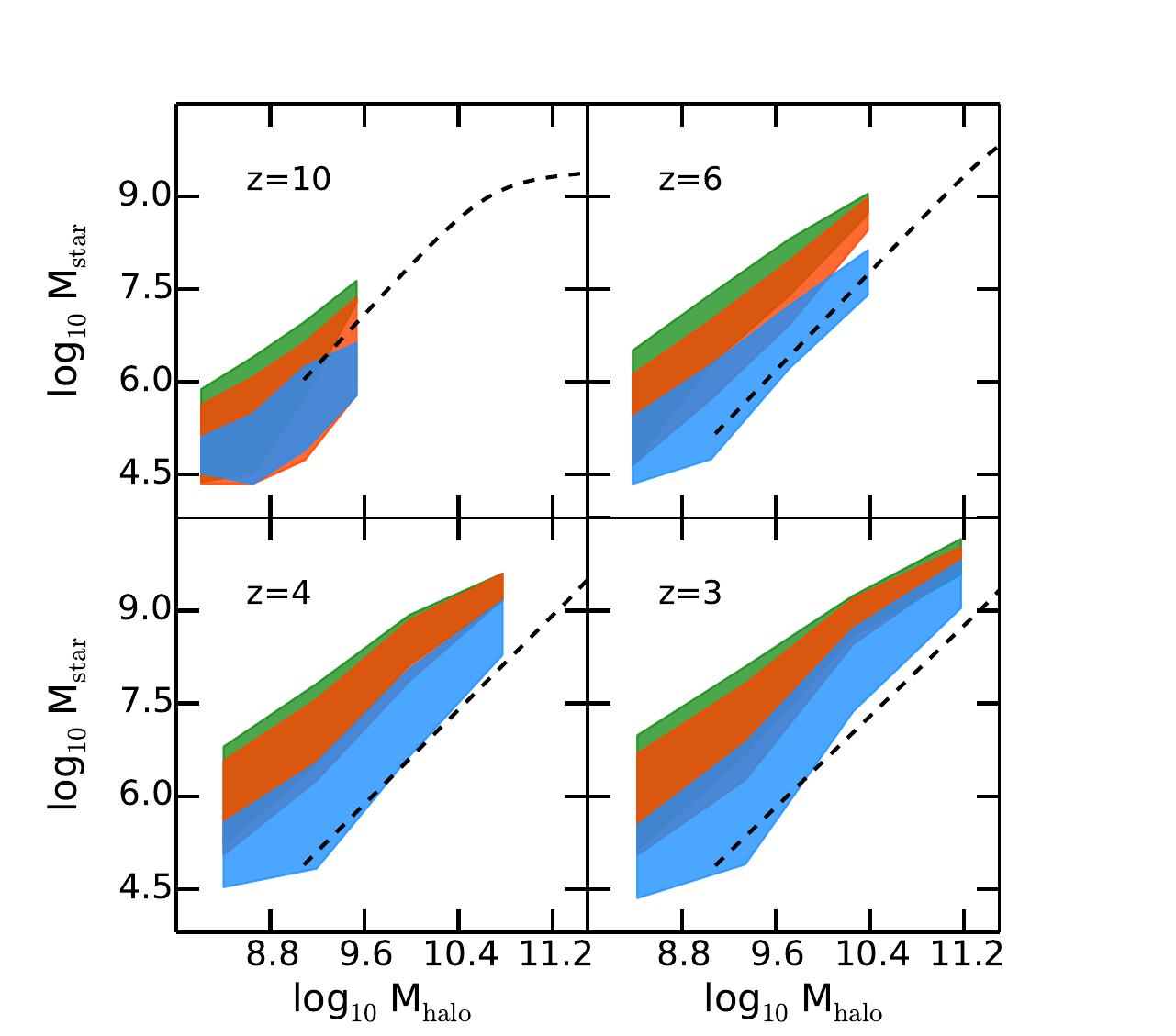}
   \caption{Stellar-halo mass relation for the  SuperChunky simulations, in green with thermal SN feedback (simulation T), in orange with the kinetic SN feedback (simulation K), and in blue with delayed cooling SN feedback (simulation D).    We show with black dashed lines an extrapolation of the empirical relation between stellar and halo masses \citep{Behroozi+13}. Simulations with thermal and kinetic SN feedbacks overestimate the stellar mass in haloes, while delayed cooling better reproduces the empirical relation.}
\label{fig:Mstar_Mhalo}
\end{figure}

\section{The influence of star formation and metallicity on BH formation}

In order to test whether star formation is realistic we compare our simulations results to the extrapolation of the model by \cite{Behroozi+13},  an empirical model of galaxy mass versus halo mass and redshift that extends up to high redshifts ($z=8$). We extrapolate to lower stellar masses and to even higher redshifts and compare it to our simulation SuperChunky.  Fig.~\ref{fig:Mstar_Mhalo} shows the stellar-halo mass relation for the T (thermal feedback, in green), K (kinetic feedback, in orange), and D (delayed cooling, in blue) simulations, with the empirical relation of \cite{Behroozi+13} as shown with dashed curves. We obtain good agreement, particularly with the delayed cooling SN feedback. 
With the thermal and kinetic SN feedbacks the stellar mass in haloes is higher, and overestimated compared to (extrapolation of) the stellar-halo mass relation. The delayed cooling SN feedback shows a better agreement of the simulation with the empirical relation.

\begin{figure}
	\centering
	\includegraphics[scale=0.47]{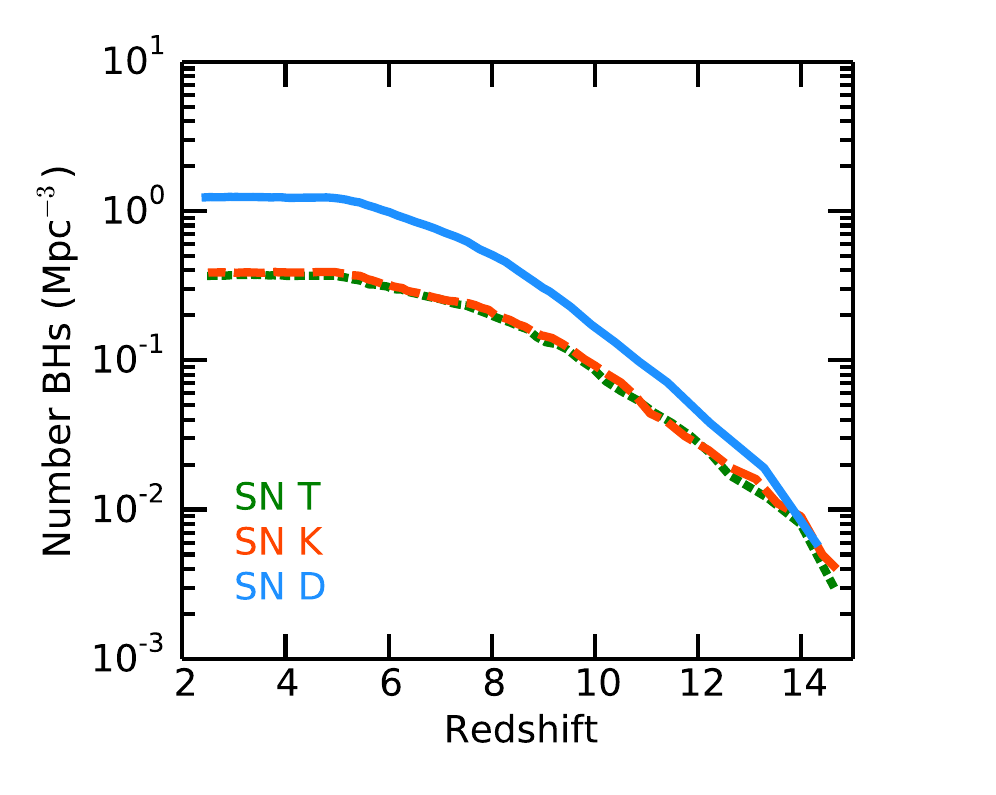}
	\caption{Number of BHs formed in the three simulations, in green for the thermal simulation, in orange for the kinetic one, and in blue for the delayed cooling one. More stars are formed in the T and K simulation, thus raising the gas metallicity.  More cold, pristine gas is still available in the D simulation to form BHs.}
	\label{fig:nb_sinks}
\end{figure} 
\begin{figure}
	\centering
	\includegraphics[scale=0.5]{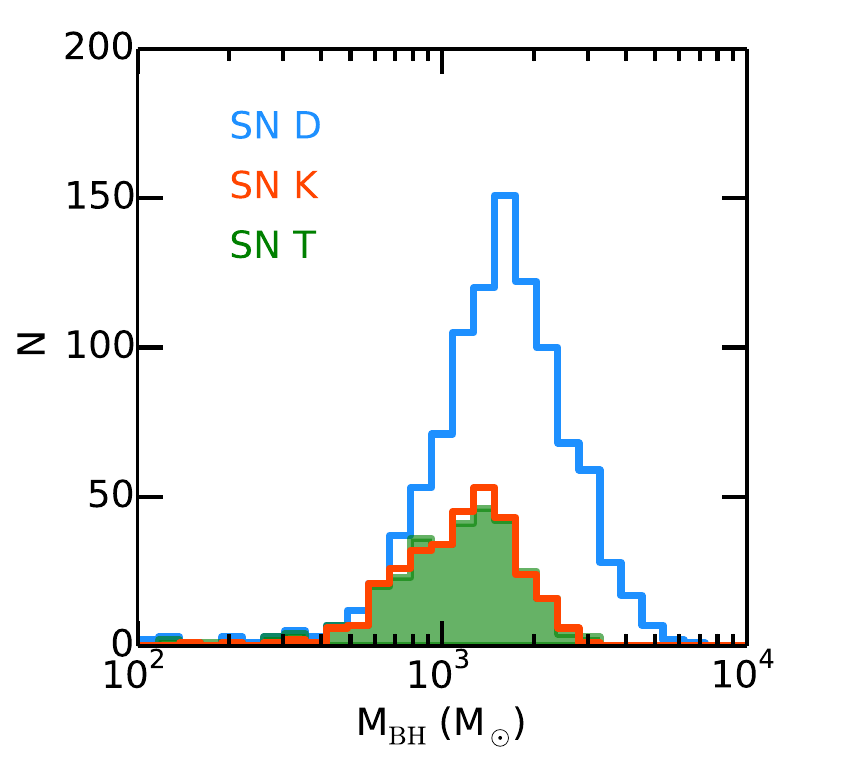}
	\caption{Initial mass function of BHs for the D (blue), K (orange), and T (green) simulations. The initial mass function of the T and K simulations are very similar. The D simulation leads to the formation of more and more massive BHs.}
	\label{fig:imf}
\end{figure}

\begin{figure*}
\centering
\includegraphics[width=\columnwidth]{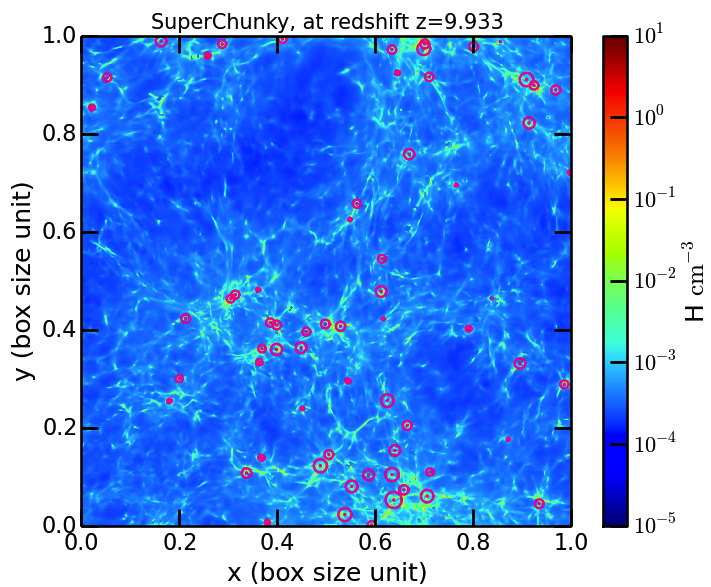}
\includegraphics[width=\columnwidth]{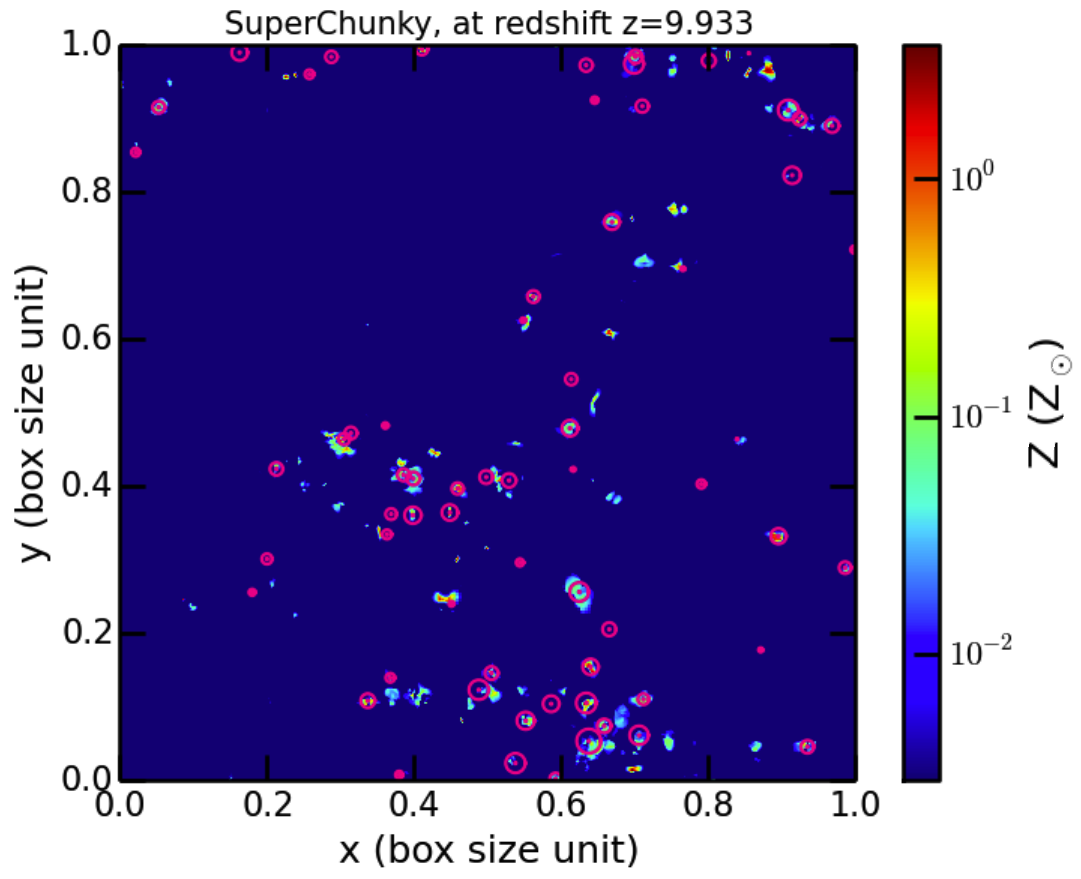}
\caption{Gas density (left, in $\rm{H/cc}$) and metallicity (right, in $Z_{\odot}$) maps at $z\sim10$ for the delayed cooling SN feedback simulation. BHs are highlighted with circles, circle sizes are proportional to BH masses. BH formation walks a fine line: the necessity of having high gas densities selects biased regions, the criterion of low-metallicity works in the opposite way. BHs form in biased regions {just before} they experience a star formation event.}
\label{fig:rho_Z_sinks_map}
\end{figure*}

More star particles are formed in the T simulation, compared to the K and D simulations. Conversely, more BHs are formed in the D simulation than in the T  and K ones. Fig.~\ref{fig:nb_sinks} shows the number of BHs formed in the three simulations over time, this correspond to the total number of sink particles at a given time in the simulation. Sink particles do not always form in the centre of galaxies and dark matter haloes, and the dynamical evolution, specifically merging and stripping, causes some of the BHs to stray into the outskirts of galaxies.  In Fig.~\ref{fig:nb_sinks}, we include all  BHs that form in the simulations, however in the following sections, we will only consider BHs within the virial radius of galaxies.

Three main features are identified in Fig.~\ref{fig:nb_sinks}, the difference in the number of sink particles in the D simulation versus  T and K, the fact that the number of BHs formed for kinetic and thermal SN feedbacks is almost identical, and the asymptotic behaviour at decreasing redshift.
In the T and K simulation, more stars are formed, therefore less cold, low-metallicity gas remains available to form a BH.  
Delayed cooling feedback is stronger, but as less stars are formed within the simulation box, the mean metallicity of haloes is always lower than in the two other simulations, and at all halo masses. The metallicity is highest in the T simulation, and K simulation is intermediate.
Regarding the similarity of the curves for the T and K cases, less stars are formed in the K simulation, therefore a larger amount of gas is available to form BHs, but the mean metal enrichment in haloes is very similar in the K and T simulations at the low-mass end. The gas is more metal-enriched, therefore the number of BHs that formed is reduced, and ends up being similar to the T simulation. The metal enrichment of the T and D simulations is discussed in \citet{2016arXiv160100557H} (Fig. 2).
Finally, at lower redshifts, the number of forming BHs tapers off; this is due to the metallicity criterion to form BHs, after $z=6$ the metal-enrichment of the medium is too large to keep forming many BHs \citep{Bellovary2011}. The three simulations follow the same trend with the saturation in the formation of pop III seed BHs below $z<5$.

 Fig.~\ref{fig:imf} shows the initial mass function of BHs which form before $z=6$, for the three simulations. The initial mass function of the T and K simulations are very similar. The D simulation leads to the formation of more BHs, these BHs are also more massive. Most BHs have masses $\sim10^3 \msun$ at birth.

Fig.~\ref{fig:rho_Z_sinks_map} shows a gas density and metallicity map at $z=10$. BHs mainly form in haloes at the intersection of filaments, i.e. in positively biased regions. This is because high gas densities are required to form and identify a gas clump in the first place. The metallicity criterion, however, acts in the opposite sense, as star formation and metal enrichment occur also in the most biased regions first. BH formation, therefore, mainly occurs  in biased regions just before widespread star formation takes place. There are pockets of metals without BHs, these are cases where no sufficiently dense clumps formed before star formation and metal enrichment made the region unsuitable for BH formation.

\section{Black hole occupation fraction}

We now turn to analyzing the distribution of BHs in halos and galaxies.  Because BHs form in dense regions, but are not forced to stay in the inner part of haloes and galaxies, we need to assign BHs and haloes, and BHs and galaxies. We consider a BH as the central BH of a halo, if its position is within 10\% of the halo virial radius. If several BHs are located within this region, we choose the most massive as the central BH. For galaxies, we proceed in the same way, looking for BHs inside the virial radius of galaxies (with a lower limit of 4 $\times$ $\Delta x$). This is the convention we will use throughout the rest of the paper. 

BH formation models do not necessarily place a BH in each and every galaxy.   One of the diagnostics to distinguish between BH formation scenarios is in fact the probability that a galaxy or halo hosts a BH: the occupation fraction. Previous theoretical studies predict a different occupation fraction of BHs in haloes for different models \citep{volonteri2008,2010MNRAS.408.1139V,Bellovary2011,2012NatCo...3E1304G}, especially in low-mass galaxies and haloes. 

\begin{figure}
	\centering
	\includegraphics[scale=0.43]{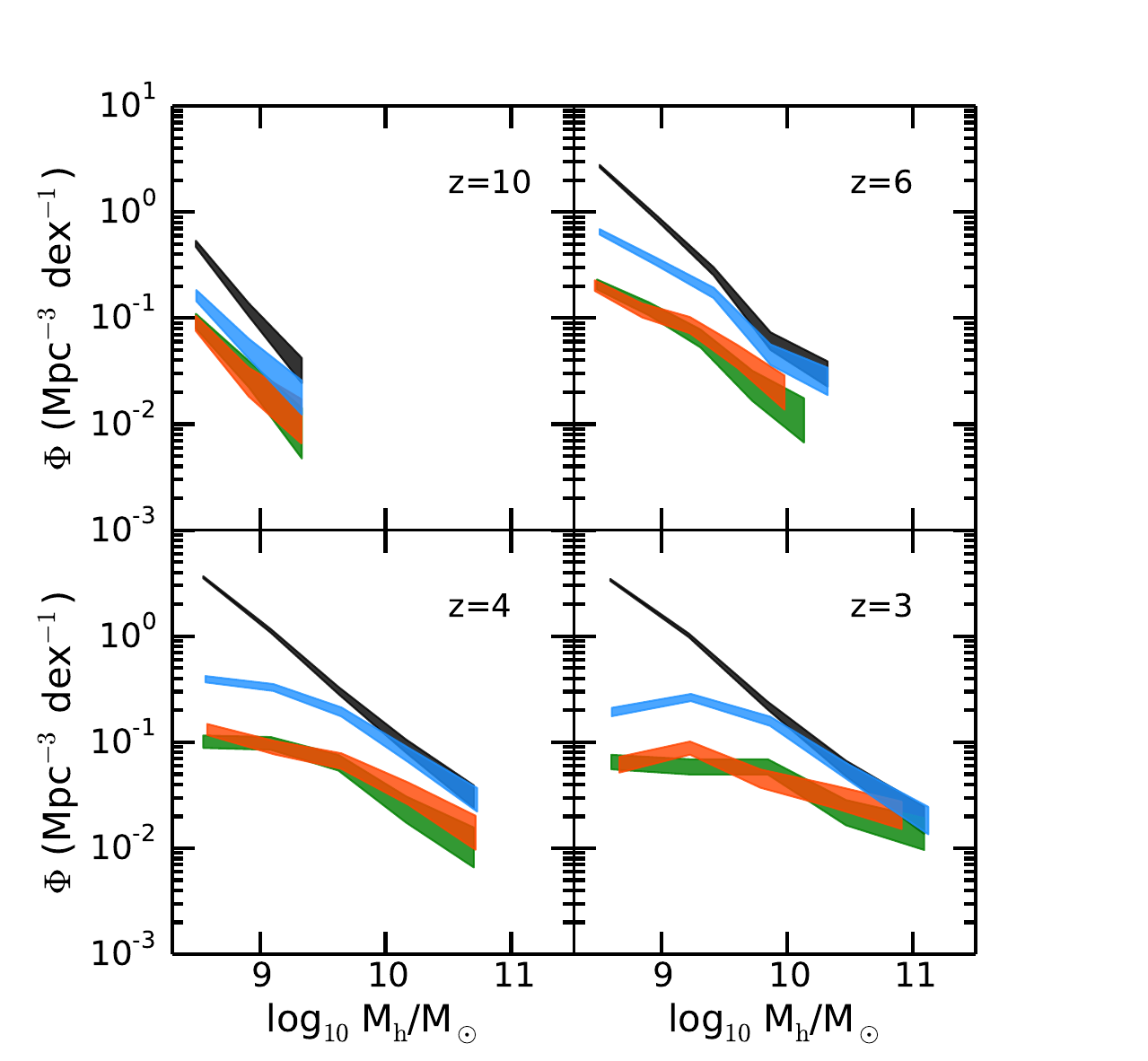}
	\caption{Total halo mass function (black; shaded areas represent poissonian error bars) with the  mass function of haloes hosting  BHs in colours (green for simulation T, orange for K, and blue for D; shaded areas also represent poissonian error bars).}
	\label{fig:BH_halo_MF}
\end{figure}

We show in Fig.~\ref{fig:BH_halo_MF} the halo mass function, where we show both the total mass function, and the mass function of haloes hosting a BH.  BH formation, indeed, does not occur in {\it all} haloes in our simulations.  Large haloes have a higher probability of hosting a BH, whereas this probability drops significantly for low-mass haloes. For the D simulation, the occupation fraction is above 50\% only for haloes more massive than $10^{9.5} \, \rm{M_{\odot}}$, or galaxies with stellar mass above $10^{7.2} \, \rm{M_{\odot}}$. For the T and K simulations, an occupation fraction of 50\% is found for higher haloes ($10^{10.5} \, \rm M_{\rm{\odot}}$) and galaxy masses (galaxies with stellar mass above $10^{8.8} \, \rm{M_{\odot}}$).
The mass function of haloes hosting a BH is closer to the total halo mass function for the delayed cooling simulation since more BHs form in this simulation.

\begin{figure}
	\centering
	\includegraphics[scale=0.53]{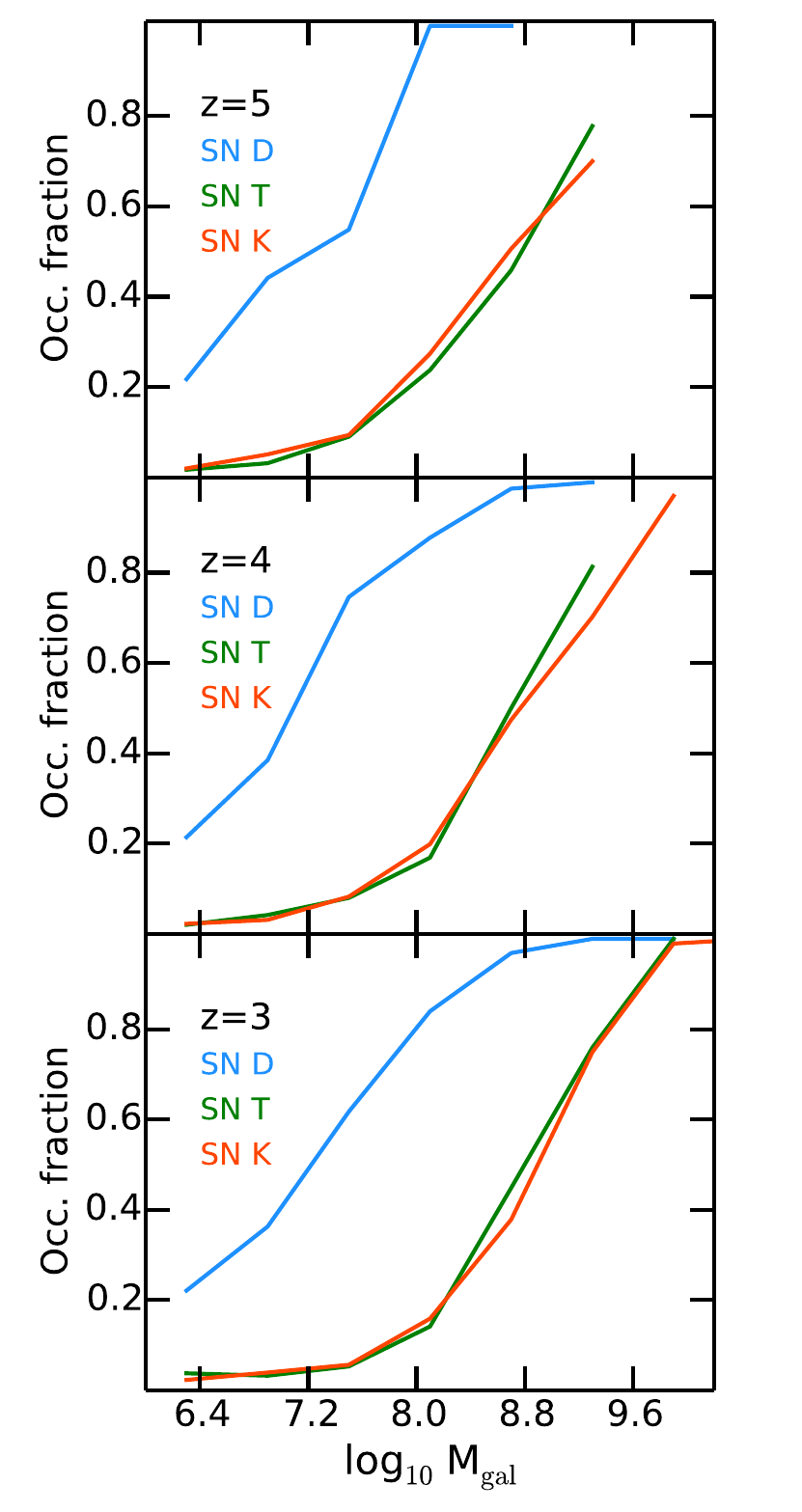}
	\caption{Probability that a galaxy of a given mass at a given redshift hosts a BH. This occupation fraction for the thermal (T,  green), and the kinetic (K, orange) SN feedbacks is similar, and lower than for the delayed cooling SN feedback (D, blue). The occupation fraction for the D simulation is higher because more BHs are formed in this simulations compared to the others, but galaxies are also less massive, because of the stronger SN feedback.  
	With time the occupation fraction shifts to the right because galaxies are becoming more massive with time, but no more BHs are formed below $z=5$. The number of BHs remains almost identical, but galaxies grow.} 
		\label{fig:galOF}
\end{figure}

We show in Fig.~\ref{fig:galOF} the BH occupation fraction as a function of galaxy mass. Here again, the probability for galaxies to host a BH is high at the high-mass end of galaxies, and drops at the low-mass end. At fixed galaxy mass, the occupation fraction is higher for the simulation with the delayed cooling SN feedback, for two reasons. First, because it allows the formation of more BHs, and second, because the strong SN feedback reduces the galaxy stellar mass, therefore the occupation fraction is shifted compared to the other simulations T and K, which have a weaker SN feedback. Regarding the evolution with redshift, the galaxy occupation fraction can increase by mergers of galaxies, and the formation of new BHs. As we have seen that no new BHs are formed since $z\sim 5$, and mergers are few, the evolution with redshift can be explained by the growth in mass of galaxies. At a given galaxy mass, the occupation fraction is lower at lower redshift, because galaxies have grown in mass. The lower redshift occupation fraction is shifted to higher galaxy masses.  This occupation fraction can be used to seed with BHs cosmological simulations at lower resolution, which can not resolve the small galaxies where BHs are expected to form. 

The fit of the BH occupation fractions (OF) can be expressed as a function depending on the stellar mass of the galaxy $x=\log_{10}({\rm M_{\star}/M_{\odot}})$ and redshift $z$. For the simulation D, we find:
\begin{eqnarray}
\begin{aligned}
&{\rm OF}_{\rm D} = 1. - \frac{0.85}{1 + (x/\varepsilon)^{\beta}}\\
&\varepsilon=-0.077 (1+z) + 7.71\\
&\beta=2.30 (1+z)^{1.32}.
\end{aligned}
\end{eqnarray}
We find the following expression for simulation T with the thermal SN feedback (a very similar behavior is found for simulation K):
\begin{eqnarray}
\begin{aligned}
&{\rm OF}_{\rm T} = 1. - \frac{0.95}{1 + (x/\varepsilon)^{\beta}}\\
&\varepsilon=-0.05 (1+z) + 9.0\\
&\beta=4.16 (1+z)^{1.1}.
\end{aligned}
\end{eqnarray}

\begin{figure}
	\centering
	\includegraphics[scale=0.55]{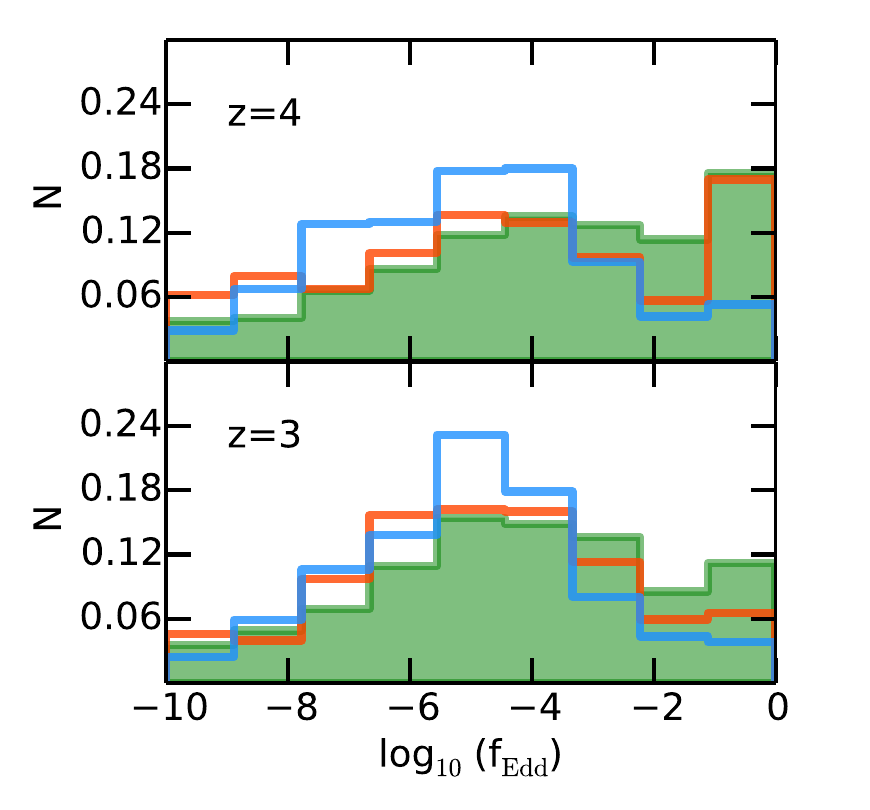}
	\caption{Normalized distribution of Eddington ratios $f_{\rm Edd}\equiv \dot{\rm{M}}_{\rm{BH}}/\dot{\rm{M}}_{\rm{Edd}}$ at redshift $z=4$ and $z=3$ for the delayed cooling (in blue), kinetic (in orange), and thermal (in green) SN feedback simulations. More BHs are accreting at the Eddington limit (when $\log_{10}(f_{\rm Edd})=0$) in the T simulation than in the D one.}
	\label{fig:fedd}
\end{figure}

\section{Black hole growth regulated by efficient SN feedback}

Most BHs in our simulations are growing slowly. The normalized histograms of the BH accretion rate in Eddington units for the three simulation are shown in Fig.~\ref{fig:fedd}. We have averaged the histogram over several outputs around $z=4$ (in the range $3.8 \leqslant z \leqslant 4.2$), and $z=3$ (in the range $2.8 \leqslant z \leqslant 3.2$). Integrated over all redshifts,  50\% of the BHs accrete at $f_{\rm Edd}\equiv \dot{\rm{M}}_{\rm{BH}}/\dot{\rm{M}}_{\rm{Edd}}<10^{-4}$ for the K and T simulation and $f_{\rm Edd}<10^{-5}$ for the D simulation. In the kinetic and thermal SN feedback simulations, a significant fraction of the BHs, 15\% are accreting at the Eddington limit, while in the D simulation the fraction of Eddington accretors in 2\%. As a consequence, BHs grow faster in the T and K  simulations than in the D simulation. 

\begin{figure*}
	\centering
	\includegraphics[width=\columnwidth]{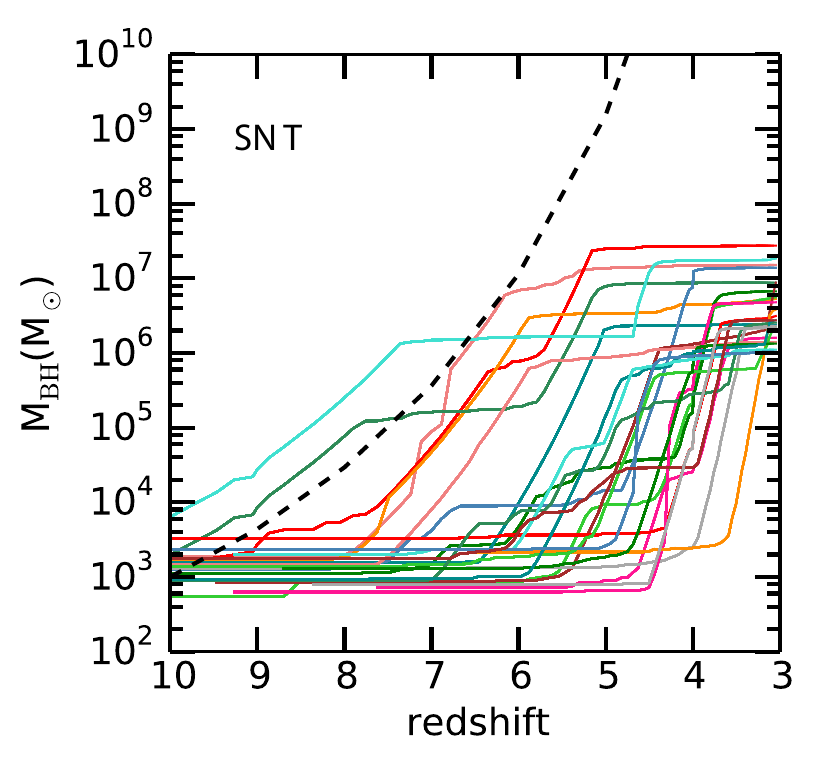}
	\includegraphics[width=\columnwidth]{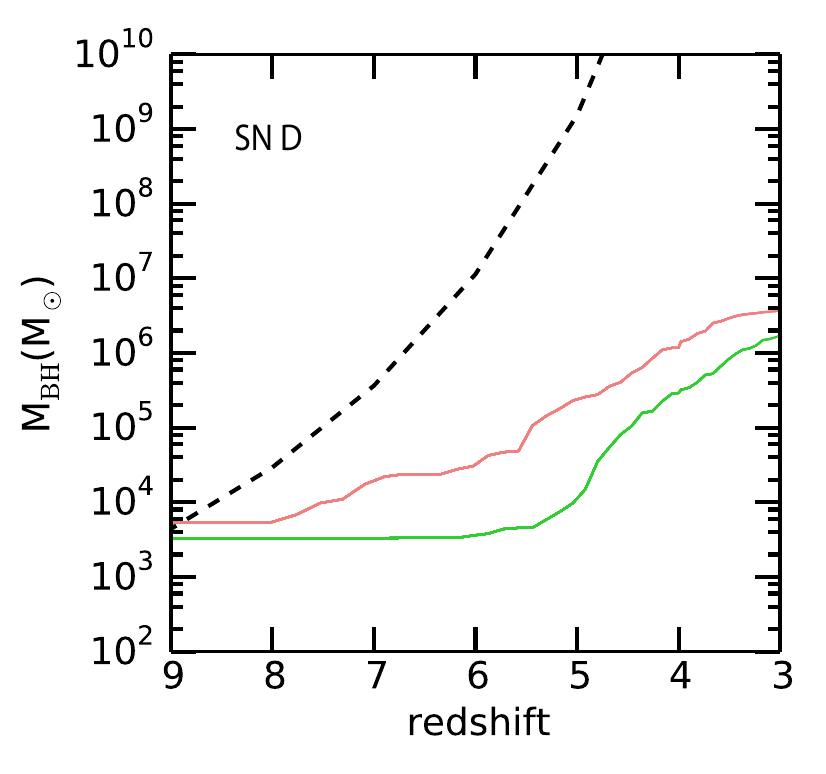}
	\caption{BH growth of the most massive BHs, $\rm{M}_{\rm{BH}}>10^{6} \rm{M}_{\rm{\odot}}$ at $z=3$, for the simulation with thermal SN feedback (left panel) and the one with the delayed cooling SN feedback (right panel). The dashed line represents BH growth at the Eddington limit. With the thermal SN feedback, BHs grow rapidly with several episodes at the Eddington limit, whereas with the delayed cooling SN feedback implementation, BH growth is smoother and delayed due to regulation of the dense gas by strong SN feedback in low-mass galaxies~\citep{2015MNRAS.452.1502D}. Only two BHs in this simulation succeed in growing to $10^{6} {\rm M_{\odot}}$ by $z=3$. We do not show the corresponding plot for the simulation with kinetic SN feedback; it is very similar to the thermal SN feedback simulation. } 
	\label{fig:BH_growth}
\end{figure*}

To investigate  BH growth in more detail, we track the BHs with mass above $10^{6}\, \rm{M_{\rm{\odot}}}$  at $z=3$. The thermal SN feedback simulation has 24 BHs above this mass threshold, the kinetic one 22, and the delayed cooling one only 2. In Fig.~\ref{fig:BH_growth}, we show the growth of these BHs with solid lines (left panel for the simulation with the thermal SN feedback, right panel for the delayed cooling one), the theoretical evolution of a BH at the Eddington limit is also showed with a dashed line. All BHs in the thermal and kinetic SN feedback simulations have episodes of accretion at the Eddington limit. In contrast, the growth of BHs in the delayed cooling simulation is smoother, and Eddington-limited phases minimal in particular at early times.

SN feedback appears therefore crucial in regulating not only star formation in low-mass galaxies, but also BH accretion. This mechanism was identified in \citet{2015MNRAS.452.1502D} as the \emph{SN-regulated growth of BHs in low-mass galaxies} by means of one single zoom cosmological simulation of group progenitor at $z=2$. We, here, confirm the result with a statistical sample of high redshift galaxies. 

In the weaker kinetic and thermal feedbacks, the energy released by SN explosions is distributed in the nearby surrounding cells, but the cooling times in dense gas cells is very short,  therefore, the cold gas present in dense central regions is not destroyed and is still available to form stars, and also to fuel a BH. This cold gas reservoir fuels the BH efficiently, and accretion occurs at rates close to the Eddington limit. In contrast, with delayed cooling SN feedback, the dense gas clumps in star-forming regions are destroyed by the release of energy after a SN explosion with a SN wind velocity of around $270\, \rm km\, s^{-1}$. The main effect is to reduce the BH growth (and star formation) in the central regions of the galaxy by routinely removing the dense star-forming gas with SN winds, until the gravitational potential well of the bulge and galaxy is deep enough to confine the cold gas close to the BH. 

\begin{figure}
	\centering
	\includegraphics[scale=0.4]{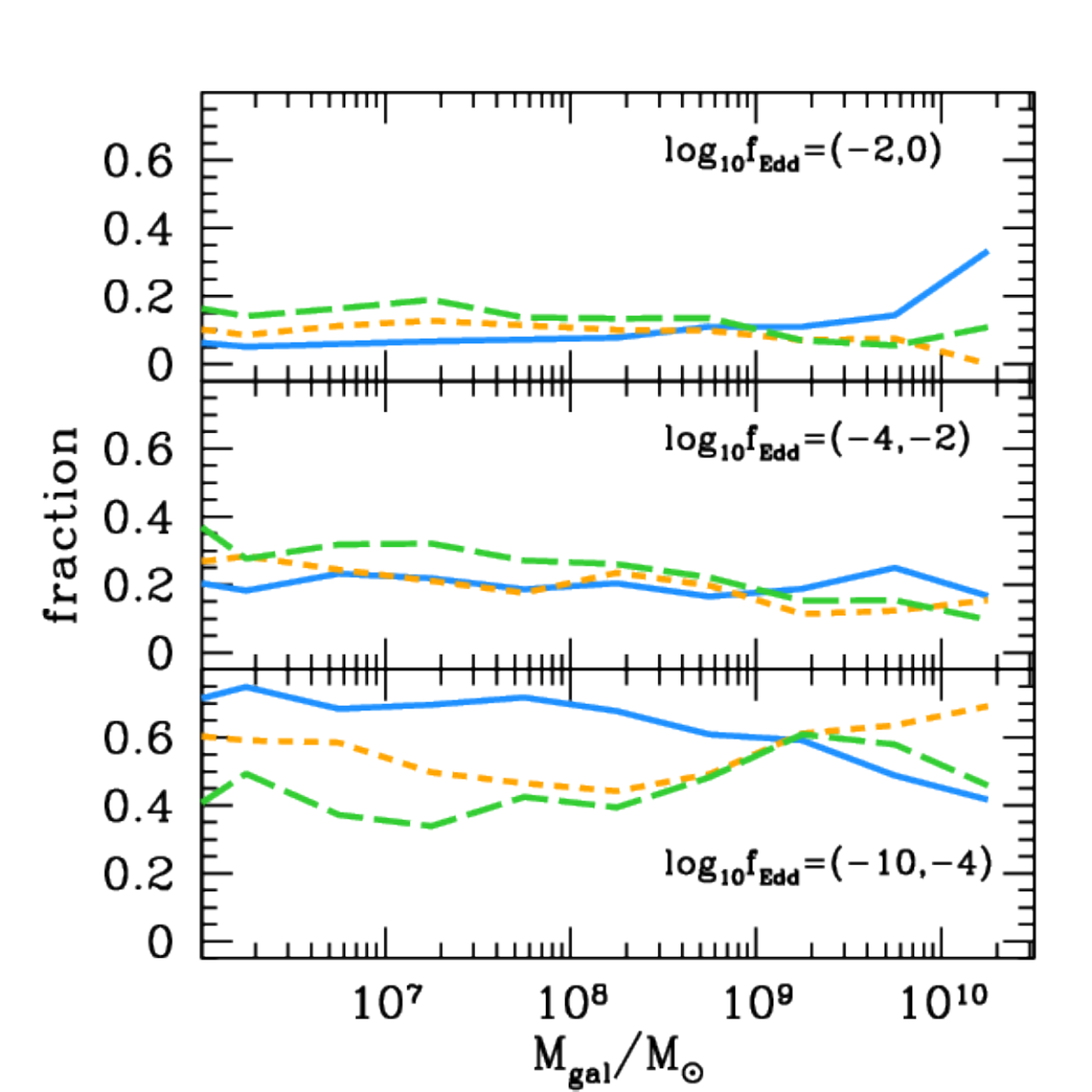}
	\caption{Fraction of BHs, at all redshifts, accreting in different Eddington ratio ranges as a function of galaxy mass. The solid  curves refer to simulation D, the dashed curves to simulation K and the dotted curves to simulation T. }
	\label{fig:duty}
\end{figure}

\citet{2015MNRAS.452.1502D} show that the growth of the rest of the galaxy is not much impacted (though growing at a slower rate than with inefficient SN feedback), therefore the total stellar mass keeps growing but not the BH mass, nor the bulge mass. It is only when the galaxy mass and bulge stellar mass become massive enough (i.e. with a corresponding escape velocity larger than SN-wind velocity) that the BH can proceed to a rapid near-Eddington growth only altered by the self-regulation due to AGN feedback. They estimated, through analytical arguments confirmed by the numerical experiment,  that this transition occurred for a galaxy stellar mass of $10^9-10^{10}\, \rm M_\odot$ ($v_{\rm esc}=300-400\, \rm km\, s^{-1}$). 
The dependence of the accretion rate, in Eddington units, from galaxy mass as a function of SN feedback can be appreciated in Fig.~\ref{fig:duty}.  In the D simulation BHs start accreting for a substantial fraction of time at high levels, $f_{\rm Edd}>10^{-2}$, only in galaxies with stellar mass $\sim 10^9-10^{10} \msun$.

It is clear that it is SN feedback, rather than AGN feedback, that determines the strength of star formation and BH growth in low-mass galaxies with low-mass BHs. In the T and K simulations, although the BHs are accreting more, and therefore dumping more energy in the host galaxy, the growth of stellar mass is more efficient, as seen e.g., in Fig.~\ref{fig:Mstar_Mhalo}, and the BH accretion rates are also higher (Fig.~\ref{fig:fedd}), implying a more effective BH growth. 

\citet{2017MNRAS.465...32B} study the build-up of the red (massive galaxies of $\geqslant$ few $10^{10} M_{\odot}$, quiescent) and blue (less massive galaxies, star-forming) sequence. They develop an analytic model, similar to the one presented by \citet{2015MNRAS.452.1502D}, where SN feedback regulates stellar and black hole growth in low-mass galaxies. Their results are generally in agreement with our simulation, although their numerical validation through the EAGLE simulation does not follow BH and galaxy growth to masses as low as ours. \citet{2017arXiv170106172P} perform a detailed comparison between the influence and interaction of SN and AGN feedback, aimed specifically at low mass galaxies in the high-redshift Universe. This study in complementary to ours in that we analyse the effect of feedback on a population of BHs, while \citet{2017arXiv170106172P} focus on the physical interactions, but on only one galaxy in a zoom.

\subsection{The assembly of black holes and galaxies}

\begin{figure*}
	\centering
	\includegraphics[width=\columnwidth]{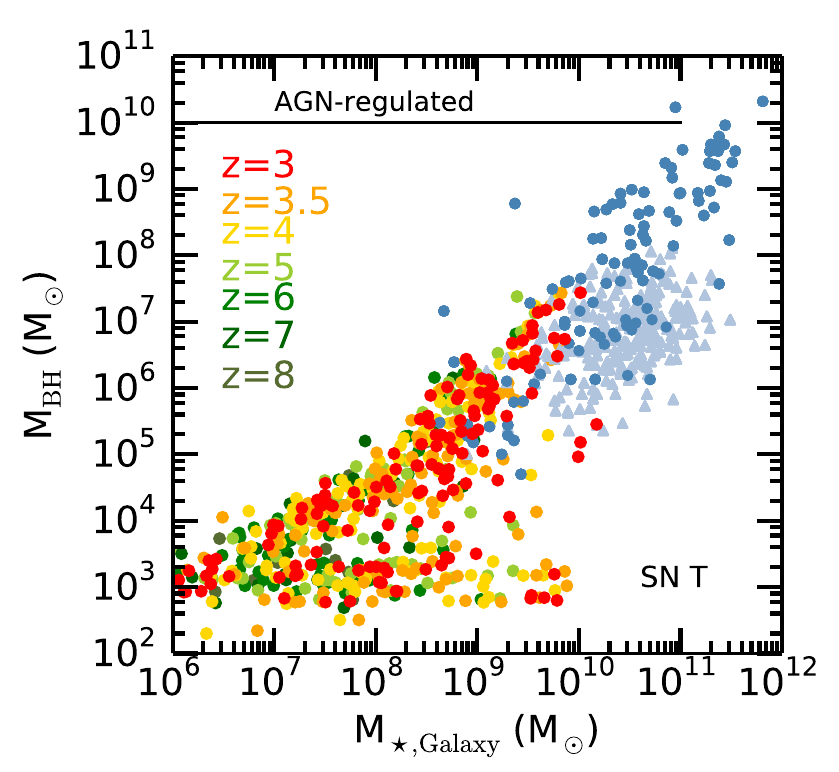}
	\includegraphics[width=\columnwidth]{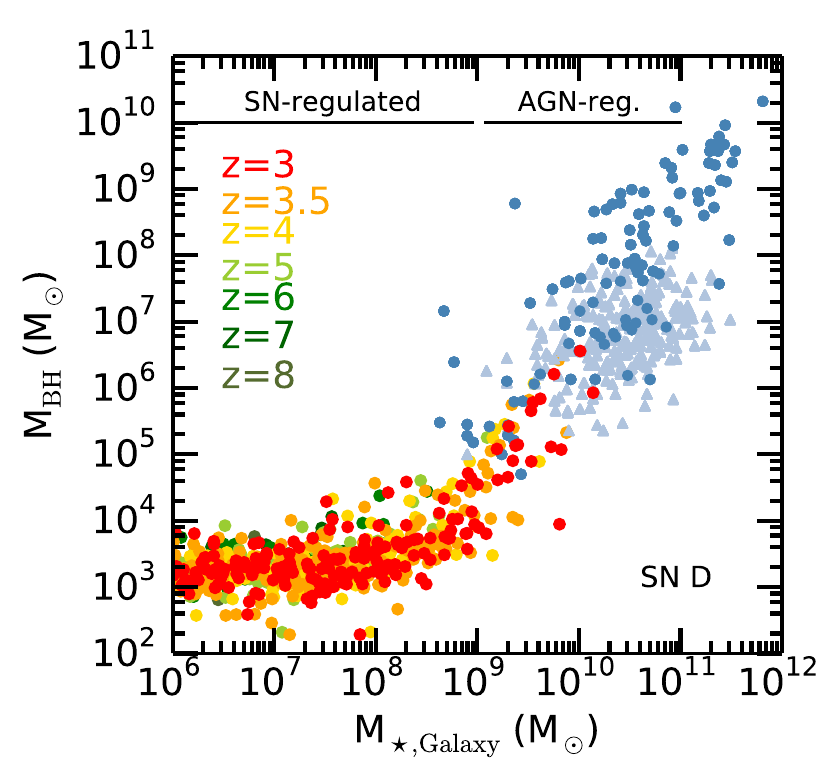}
	\caption{BH mass as a function of the total galaxy stellar mass for the thermal SN feedback (left) and the delayed cooling SN feedback (right) together with observations from \citet{2015arXiv150806274R} (blue points). In the ``strong'' SN feedback case (D), BHs in low-mass galaxies ($\rm{M_*}<10^9\, \rm{M_\odot}$) have a hard time to grow because these galaxies have shallow potential wells as a result of SN winds, which are sufficient to remove the dense star-forming gas and suppress BH accretion~\citep{2015MNRAS.452.1502D}.}
	\label{fig:MBH_Mgal}
\end{figure*}

Fig.~\ref{fig:MBH_Mgal} shows the BH mass as a function of the total stellar galaxy mass, at several redshifts ($z=8,7,6,5,4,3.5,3$), for the thermal SN feedback simulation on the left panel and the delayed cooling one on the right panel. The BH-galaxy mass relation for the kinetic feedback simulation (not shown here) is very similar to that of the thermal feedback case.  

In the thermal SN feedback simulation, some massive galaxies host very low-mass BHs. These are BHs that have recently been acquired from a satellite galaxy that merged with a larger galaxy that did not initially host a BH. We do not force BHs to form in massive galaxies, in fact, if in some galaxy wide-spread star formation and metal enrichment occur before the formation of a dense, bound clump that meets all the criteria for BH formation, that galaxy is not seeded with a BH. Fewer BHs form in the thermal and kinetic SN feedback case, as discussed in Section~4.1, therefore more galaxies are BH-less.  Some BHs form in relatively small galaxies, which travel to intersections of filaments, and are then captured by a more massive galaxies. If the more massive galaxy does not host its own BH, eventually the BH in the satellite galaxy can become the central BH of the merger remnant. We do not reposition BHs at galaxy centres artificially, but let them evolve under the effect of dynamical friction \citep[see][for a detailed discussion on BH dynamics in cosmological simulations]{2015MNRAS.451.1868T}. The timescale for a small BH ($\sim 10^3\, \rm M_\odot$) to settle in the galaxy centre is long, of order of a few hundred Myr to Gyr  \citep{binney1987}, and, during the orbital decay, the rapidly moving BH cannot efficiently accrete gas from its surroundings. Only after the BH has settled long enough in the galaxy centre, it will start accreting and grow, ``catching up" with its galaxy. Such population is instead not present in the delayed cooling feedback simulation, simply because a larger fraction of galaxy is initially seeded with their own BH: the BHs initially hosted in satellite galaxies either merge with the pre-existing BH in the main galaxy, or remain stranded in its outskirts without necessarily merging with the central BH \citep[see, e.g.,][]{Islam2003,VolonteriPerna2005,2016arXiv160201941V}.

Except for the small population of recently captured BHs, we see that ``weak'' SN feedback (T and K) produces a near-linear BH-galaxy mass relation, while for the ``strong'' SN feedback (D) the BH-galaxy mass relation plateaus at low galaxy mass with an ankle at $\rm M_{\star}=10^9-10^{10}\, \rm M_\odot$ and a steep rise above this mass transition. 
From Fig.~\ref{fig:MBH_Mgal}, we see that BHs are indeed growing faster in the T and K simulations than in the D simulation, and as a result the BH masses are larger relative to their galaxy stellar mass in the T and K simulations. As we have explained before, this is due to the SN feedback being weaker in the T and K simulations, leaving cold gas available in the central region of the galaxy to fuel a BH. Whereas in the D simulation, SN feedback is stronger, the cold gas of the central region is removed and BH growth is therefore reduced.

In Fig.~\ref{fig:MBH_Mgal} we report also the BH and stellar mass for the objects published by \citet{2015arXiv150806274R}. In this paper, 262 broad-line AGN and 79 galaxies with dynamical BH mass measurement, for redshift $z<0.055$, are used to investigate the scaling relation between BH mass and the total stellar mass of galaxies. We see that when BHs grow, they  eventually connect to the low-redshift sample. In low-mass galaxies, however, BHs are unable to grow, and more so if SN feedback is strong, and BHs remain ``stuck" at low mass.   \citet{2011MNRAS.417.2085V} proposed, based on empirical arguments, that if  BHs in small galaxies are under-massive and BHs in large galaxies are over-massive, then one can reconcile several observational results, namely that analysis of the BH mass/luminosity function and clustering suggests that either many massive galaxies do not have BHs, or these BHs are less massive than expected \citep{Willott2010,2013ApJ...778..130T,2015MNRAS.448.3167W}.

In Fig.~\ref{fig:MBH_Mgal}, horizontal lines show the regulation phases of BH growth, either driven by SN feedback or by the AGN itself. In the simulations T and K, because SN feedback is weak, BHs can rapidly grow at early times, and quickly regulate their own growth, they are {\it AGN-regulated}. Conversely, in the delayed cooling model, SN feedback is stronger, which leads to a slower growth of the BHs, and their accretion is highly sub-Eddington (see Fig.~\ref{fig:fedd}). The early growth of BHs is inhibited by SN feedback, and therefore BHs are in a {\it SN-regulated} phase. Eventually, when enough cold gas has been accreted again in the central part of galaxies to fuel BHs, and the central potential well is deep enough to retain this gas against SN feedback, they enter the {\it AGN-regulated} phase.

\begin{figure}
	\centering
	\includegraphics[scale=0.5]{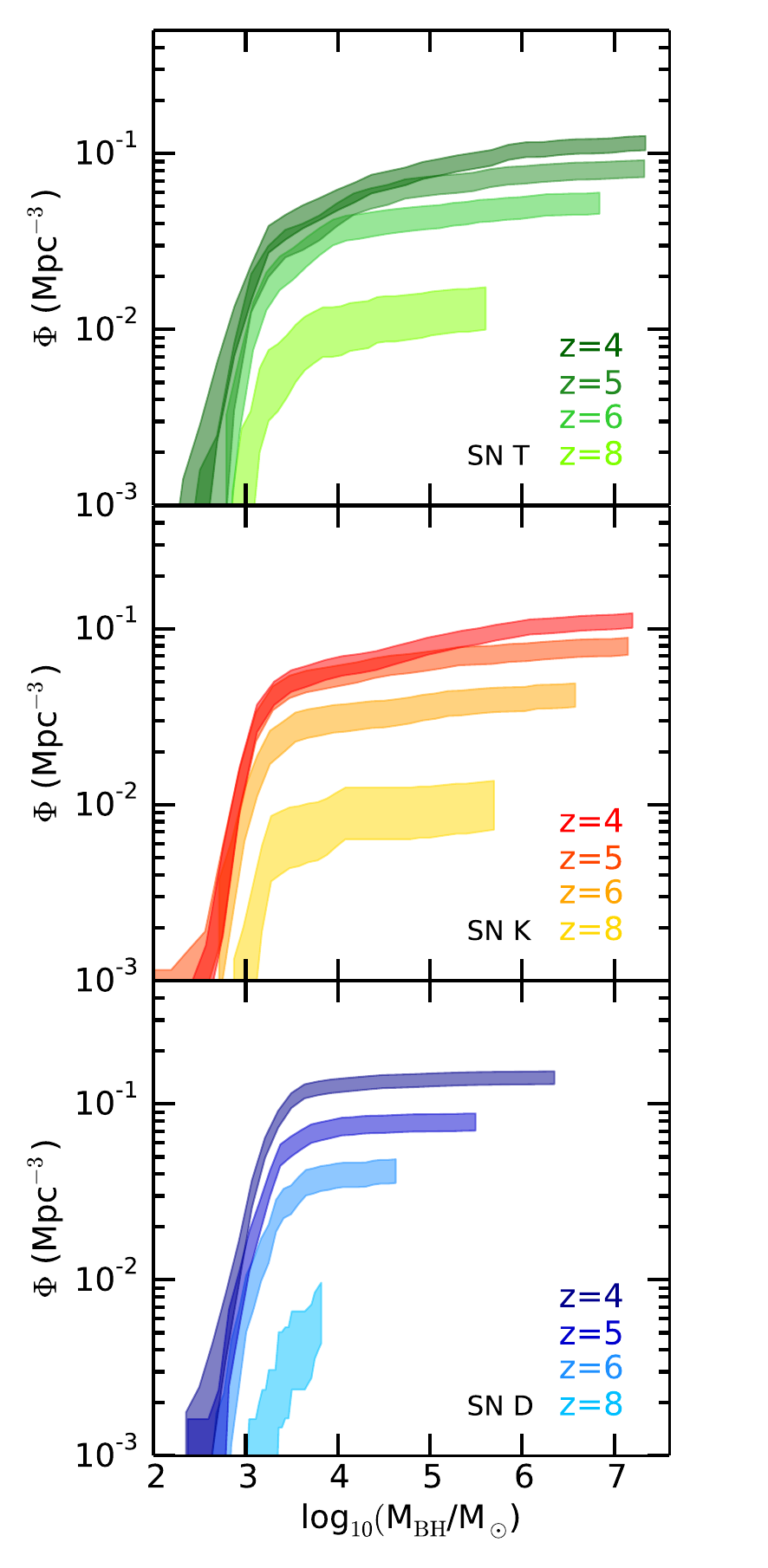}
	\caption{Cumulative BH mass function for the thermal SN feedback simulation (T, top panel), the kinetic one (K, middle panel), and for the delayed cooling one (D, bottom panel), for redshift $z=8,6,5,4$. Less BHs are produced in the T and K simulations, but they grow more, up to $\sim 10^{7} \rm{M_{\odot}}$, about one order of magnitude higher than for the delayed cooling one, which is a signature of SN feedback regulating the growth of low mass BHs in the case of strong feedback (delayed cooling).}
	\label{fig:BH_MF}
\end{figure}

\subsection{Black hole mass function}

 Fig.~\ref{fig:BH_MF} shows the cumulative mass function of BHs at different redshifts. This differs from the initial mass function of BHs (Fig.~\ref{fig:imf}), as we take into account both the seed mass, the mass accreted by the BHs and BH-BH mergers (which are sub-dominant in the mass growth budget). We show in Fig.~\ref{fig:BH_MF} the three different simulations, with the thermal, kinetic, and the delayed cooling SN feedbacks. 

The evolution with time is as expected: with increasing cosmic time (decreasing redshift), more and more BHs form, and the already formed BHs grow in mass. Although the mass functions are very similar for the T and K simulations, the low-mass and high-mass ends slightly differ. More central BHs are identified in the kinetic feedback simulation. The high-mass end of the distribution is higher in the T case, showing that the weak thermal SN feedback favours the growth of BHs. The delayed cooling simulation has the largest number of BHs. Conversely, BHs do not grow in mass as efficiently as in the T and K simulations. The strong SN feedback limits the growth of BHs.  In summary, BHs are more numerous in the delayed cooling SN feedback simulation, but their masses are smaller. Only few BHs reach a final mass of $10^6\, \rm{M_{\odot}}$ by $z=3$ with a strong SN feedback. In the T simulation, because the SN feedback is weaker, BHs can grow faster to even higher mass (several BHs reach $10^7\, \rm{M_{\odot}}$ by $z=3$).

\section{Comparisons with observations}

\begin{figure}
	\centering
	\includegraphics[scale=0.344]{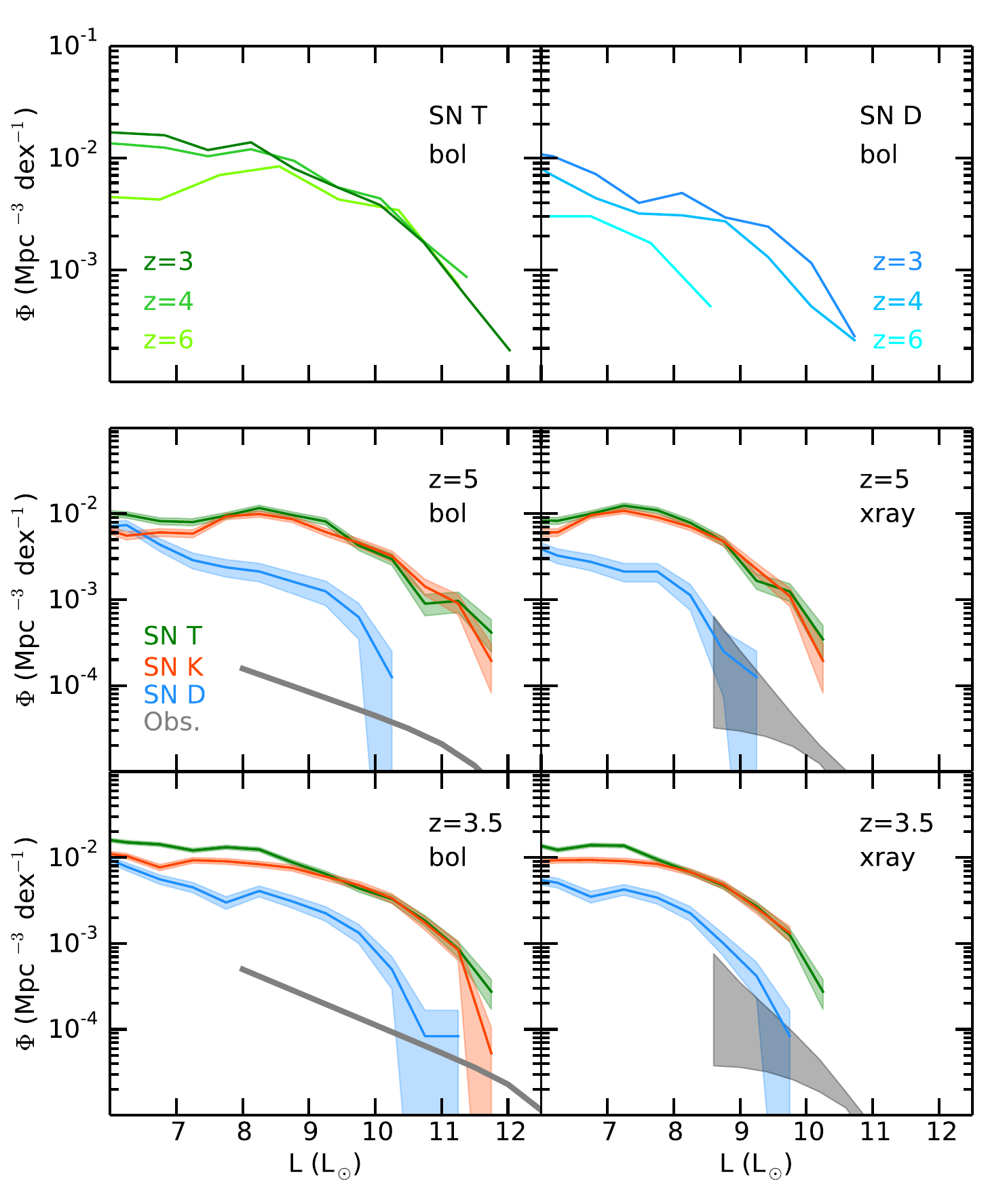}
	\caption{Luminosity function of simulated BHs, and comparison with observational constraints. In the two top panels, we show the evolution of the bolometric luminosity function of the simulated BHs with time (for $z=6,4,3$) for the simulation T (top left panel) and the simulation D (top right panel). In the 4 bottom panels, we compare the bolometric (left) and the hard X-ray (right) luminosity function of  simulated BHs in simulations D (blue), K (orange), T (green) with observations \citep[shaded grey regions,][]{Hop_bol_2007,2015ApJ...802...89B}. The middle panels show the comparison at $z=5$, and bottom panels at $z=3.5$.}
	\label{fig:plot_luminosity_comp_hopkins_obs}
\end{figure}

\subsection{High redshift AGN}
Over the past decades, an incredible effort has been made to study, theoretically and observationally, the BH, quasar and AGN population in massive galaxies, very often looking for the smoking gun of AGN feedback, thought to impacting the growth and star formation at the high-mass end of the galaxy distribution. In Fig.~\ref{fig:plot_luminosity_comp_hopkins_obs} we present the bolometric luminosity function of BHs at different redshifts, for simulation T (top panel on the left) and simulation D (top panel on the right). The bolometric luminosity of the simulated BHs, $\rm{L}_{\rm{bol},\odot}$, expressed in solar luminosity, is defined by:
\begin{eqnarray}
\rm{{L}_{\rm{bol},\odot}}=\epsilon_{\rm r} \, \dot{\rm M}_{\rm acc} c^{2} -33.6,
\end{eqnarray}
with the radiative efficiency $\epsilon_{\rm r}=0.1$.
The luminosity function from the simulation T (and similarly simulation K) is higher than the one from the simulation D, and also pushed to higher luminosities, because with the weaker SN feedback BHs in the simulations T and K are accreting at higher rates.  
We also compare the luminosity function of our simulated BHs to the fit to the bolometric luminosity function compiled by \cite{Hop_bol_2007} and its extrapolation (middle and bottom panels on the left in Fig.~\ref{fig:plot_luminosity_comp_hopkins_obs}), and to the X-ray luminosity \citep[middle and bottom panels on the right,][]{2015ApJ...802...89B}.  
The hard X-ray luminosity of the simulated BHs is computed by applying a bolometric correction $\rm BC$, following \citet{Hop_bol_2007}:
\begin{eqnarray}
\log_{10}\left(\rm{L}_{2-10\, \rm{keV},\odot}\right)=\log_{10}\left(\rm{L}_{\rm{bol},\odot}\right)-\log_{10}(\rm BC),
\end{eqnarray}
\begin{eqnarray}
\rm{BC}=10.83 \left(\frac{\rm{L}_{\rm{bol},\odot}}{10^{10}\,\rm{L_{\odot}}}\right)^{0.28} + 6.08 \left(\frac{\rm{L}_{\rm{bol},\odot}}{10^{10}\, \rm{L_{\odot}}} \right)^{-0.020}.
\end{eqnarray}

Our low-mass, slowly accreting BHs are well below current observational limits. Future high-sensitivity missions, such as JWST \footnote{http://www.jwst.nasa.gov} and ATHENA \footnote{http://www.the-athena-x-ray-observatory.eu},  and proposed ones such as X-ray Surveyor \footnote{https://zenodo.org/record/47667/files/Civano-Francesca.pdf} and StarX, can instead start probing the luminosity range where normal high redshift BHs in normal galaxies are evolving, instead of the brightest quasars powered by the most massive BHs we can reach today.

The strongest observational constraints on high redshift AGN come from analyses of the central area of the CDF-S, $\sim 150$ arcmin$^2$, where the flux limit reaches a nominal value of $7\times 10^{-18}$ in cgs units in the 0.5-2 keV band, corresponding to an X-ray luminosity of  $\log({\rm L}_{\rm X})>42.2$ in cgs units at z=6 (with $2\times 10^{-17}$ and $\log({\rm L}_{\rm X})>42.9$ being more conservative values; R. Gilli private communication).  Currently, there are only 3 candidates at $z>6$ (based on  photometric redshifts, where possible AGN contamination is not taken into account) in the survey area in this redshift range \citep{2015A&A...578A..83G}, and even these sources are debated \citep{2015MNRAS.448.3167W, 2015arXiv151200510C, 2016MNRAS.463..348V}. We estimate the number of AGN we predict at $6<z<7$ in our runs in a corresponding area of the sky. In an area of the sky of $150$ arcmin$^2$ we find $\sim 5.4\times 10^3$ AGN above $\log({\rm L}_{\rm X})>42.2$ at one point between $z=6$ and $z=7$ in simulation T, $\sim 2.2\times 10^3$ in simulation K and zero in simulation D. Extrapolating the X-ray LF of simulation D to higher luminosities, we would predict $\sim$110 AGN. The typical lifetime of each AGN is 0.03 Gyr, and the cosmic time between $6<z<7$ is 0.171 Gyr, giving a duty cycle of 0.175, and bringing the expected number to $\sim 960$ in simulation T, $\sim 380$ in simulation K, and $\sim 20$ from the extrapolation of the X-ray LF  in simulation D. At these redshifts, obscuration in Compton thin sources, with column density $N_{\rm H}=10^{22}-10^{24} \rm{cm}^{2}$, should not hinder detection, but Compton thick sources, with  column density $N_{\rm H}>10^{24} \rm{cm}^{2}$ would still be missed. Such population is expected to account for 30-50\% of the AGN population, based on lower redshift hard X-ray observations \citep[][and references therein]{2014ApJ...786..104U} and synthesis of the X-ray background \citep{2007A&A...463...79G}. Such correction brings the number of predicted AGN to $\sim 480-670$ in simulation T, $\sim 190-270$ in simulation K, and  $\sim 10-13$ from the extrapolation of the XLF in simulation D. Taking the more conservative limit of $\log({\rm L}_{\rm X})>42.9$, the number of AGN, including duty cycle and Compton Thick correction, would be $\sim 220-310$ (T), $\sim 30-45$ (K) and $\sim 3-5$ (D) for the three simulations. The several tens or hundreds of AGN predicted by simulations T and K appear in contrast with current data, while the results for simulation D, though nominally higher than the current number of candidates/upper limits, are in much better agreement with current constraints.

\subsection{Low redshift analogues of high redshift galaxies}

The high-mass end of the BH distribution provides us with essential information on the growth of BHs. However, as the host galaxies are typically massive, all the clues relating to BH formation have been erased by the growth of BH, through gas accretion and BH-BH mergers \citep[e.g.,][and references therein]{2014MNRAS.440.1590D}.  In order to collect crucial information on BH formation, one has to look at the least evolved galaxies. Lacking observational samples of low-mass galaxies at high redshift, we compare our simulations to two different types of low-redshift galaxies: dwarf galaxies and local analogs of high redshift galaxies (LBAs).  

\begin{figure}
	\centering
	\includegraphics[scale=0.55]{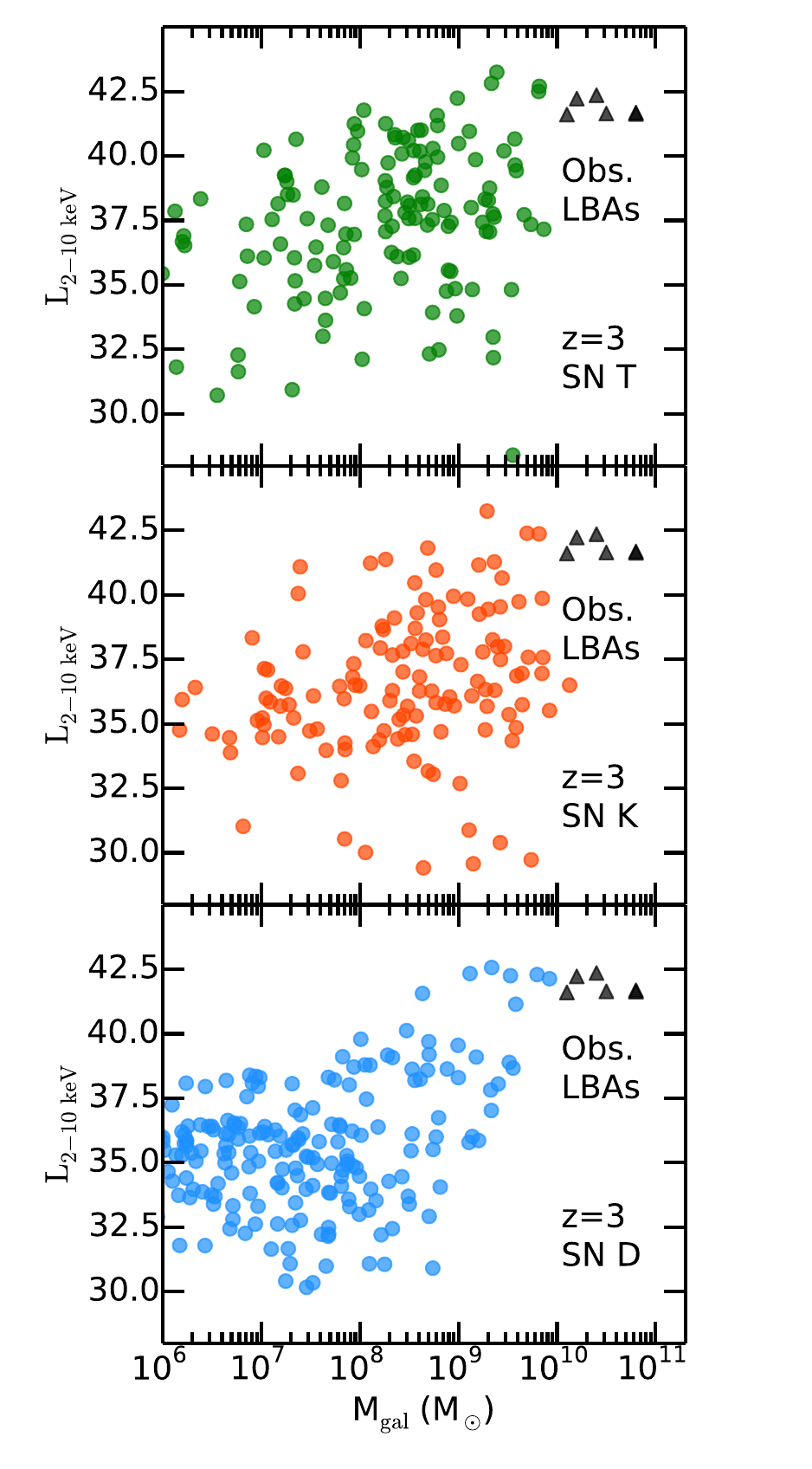}
	\caption{BH hard X-ray luminosity from the simulations (delayed cooling in blue, kinetic SN feedback in orange, thermal SN feedback in green, at redshift $z=3$), compared to the observed BHs in LBAs (black triangles). The luminosity, and so the accretion rate, of the most luminous simulated BHs is similar to the LBAs, suggesting that the physical conditions are at least comparable.}
	\label{fig:plot_LBA}
\end{figure}

 In recent years, many studies have started looking for evidence of the presence of BHs in low-mass galaxies (with stellar mass of $\rm{M_{\star}} \sim10^{9} \rm{M_{\odot}}$):  evidence for accreting BHs with broad H$\alpha$ line in the SDSS survey \citep{2004ApJ...610..722G, 2007ApJ...667..131G, 2012ApJ...755..167D}, evidence for narrow-line AGN in low stellar velocity dispersions (favoring the presence of a low mass BH), or at higher redshift with stacking methods \citep[for example, X-ray stacking up to $z=1.5$][]{2016ApJ...817...20M}.
\citet{2013ApJ...775..116R}  went further and performed the first systematic search for BHs in galaxies with stellar mass of $\rm{M_{\star}}<3 \times10^{9} \rm{M_{\odot}}$. They found 136 dwarf galaxies harboring evidences of active BHs (photoionization signatures, broad emission lines). The comparison between our high redshift samples (from $z=8$ in green to $z=3$ in red points) and the observations in the local Universe (dark and light blue points) is shown in Fig.~\ref{fig:MBH_Mgal}. We discussed how BH growth appears to be stunted in low-mass galaxies by SN winds, but with our high redshift simulations we can only extrapolate our results to the local Universe provided that the BH to galaxy mass relation show little-to-no evolution with redshift~(e.g. \citealp{2016arXiv160201941V} for the Horizon-AGN simulation,~\citealp{2014MNRAS.444.1453D}); bringing such high-resolution simulation as SuperChunky to $z=0$ is computationally very expensive.

 A fairer comparison can be made with local analogs of high redshift galaxies. Such LBAs are promising laboratories for constraining BH formation. They have properties similar to the more distant LBGs, in terms of mass, age, size, metallicity, star formation rate, optical extinction, but are much closer to us, thus permitting more detailed studies.  AGN in LBAs can then provide us crucial clues on BHs in LBGs and then directly on the high redshift population of BHs. Fig.~\ref{fig:plot_LBA} compares the sample of BHs in our simulations with the available observations in LBAs, i.e. the sample of XMM observation of six LBAs described in \citet{2011ApJ...731...55J}, with $z \leqslant 0.3$. We postpone a detailed comparison with LBAs and LBGs to a future paper, and we show here only the normalized distribution of accretion rate, shown as X-ray luminosity. The accretion rate is similar for the high redshift galaxies for the most luminous BHs in the simulation and the LBAs (shown as the black triangles in Fig.~\ref{fig:plot_LBA}) suggesting that the physical conditions are at least comparable. 

What emerges from the comparison with observations is that the delayed cooling SN feedback model is the one producing a population of simulated galaxies and BHs in better agreement with the current available observations. It better reproduces the stellar-halo mass relation of the empirical model of \citet{Behroozi+13} (Fig.~\ref{fig:Mstar_Mhalo}), the simulated BHs connect to the local galaxy sample of \citet{2015arXiv150806274R} (Fig.~\ref{fig:MBH_Mgal}),  the BH luminosity function of this model is the closest to the bolometric and X-ray luminosity functions derived respectively by \citet{Hop_bol_2007}, and \citet{2015ApJ...802...89B} (Fig.~\ref{fig:plot_luminosity_comp_hopkins_obs}), and finally the model provides a number of AGN in better agreement with the number of AGN candidates detected in the CDF-S survey \citep{2015A&A...578A..83G} at high redshift.

\subsection{Inferences on the assembly of the most massive black holes}

The population of galaxies and BHs formed in the D simulation is in better agreement with current observations, but the BHs formed in this simulation are less massive than in the T or K simulations, and exploring whether this model can also account for the most massive BHs is important. One of the main concerns  are the $z\sim 6$ quasars powered by billion solar mass BHs. Such BHs are expected to be hosted in massive halos in very biased regions, therefore they are very rare, $<1$ cGpc$^{-3}$.  Because we are using a simulation with a volume of $10^3$ cMpc$^3$, we do not have massive galaxies or halos in our simulations. Quasars at high redshift are thought to reside in very massive halos, of about $10^{13} \msun$ at $z\sim 6$, while our simulations only probe halos of  $<10^{11} \msun$ at the same redshift (cf. Fig.~\ref{fig:BH_halo_MF}). We therefore do not expect to find $10^{9} \msun$ BHs in our simulations. 

We can, however, discuss our results in light of previous/concurrent work, specifically  two types of different approaches, (i) smaller volumes, normally a zoom on a single halo, at higher resolution \citep[e.g.,][]{Alvarez2009,Johnson2011,2014ApJ...797..139A}, and  (ii) larger halos and volumes at lower resolution aimed at studying $z\sim 6$ quasars \citep[e.g.,][]{Dubois2013,2014MNRAS.439.2146C,2014MNRAS.440.1865F,2016arXiv160608871D}. Our simulations bridge the gap between these two regimes. 

In the smaller/higher-resolution simulations the typical result is that BH growth is limited to be sub-Eddington, because of AGN feedback,  {\it even} when not including SN feedback. In the presence of SN feedback the suppression of accretion is strengthened (Latif et al. in prep.). All these simulations focused on small BHs in relatively small halos, $10^7-10^8 \msun$.   In the larger/lower resolution simulations where the minimum halo mass which is resolved is a few times $10^8-10^9 \msun$, BH growth appears always to jump from basically zero to Eddington-limited at $z\sim10-13$ (see e.g., Fig. 6 in Dubois et al. 2016 and Fig. 6 in Costa et al. 2014), with a behaviour resembling those of our BHs, but shifted to higher redshift. In fact, a more massive, more biased halo would reach sufficiently high mass to allow for BH growth in the presence of SN feedback at earlier times.   The change in the gradient of BH growth at sufficiently high stellar mass can, for instance, be appreciated in Fig. 4 in di Matteo et al. 2016. Costa et al. 2014 also  comment that {\it when halos become massive enough}, processes that eject gas from halos inner regions are less efficient, and a reservoir of gas is still available to fuel central BHs. Additionally, they note that outflows from smaller galaxies around a more massive one also help growing the BH in the massive galaxy, as a fraction of the gas expelled from these small galaxies is incorporated into the more massive one, boosting its gas content.  Therefore, overall, the results of both smaller/higher-resolution and larger/lower resolution simulations bracket and validate our results, at least qualitatively. The assembly of more massive halos provide a deeper potential earlier on, helping the growth of more massive BHs, while in the fragile environment of smaller, more isolated halos, BH growth is inhibited.

\section{Conclusions}
In this paper, we present a new implementation to seed cosmological simulations with BHs. Our implementation mimics BH formation models based on stellar properties, namely a scenario based on remnants of PopIII stars, and a scenario based on stellar mergers in nuclear star clusters at low metallicity. The seed BHs are relatively small, as expected for the scenarios investigated here. Most BHs have masses $\sim 10^3 \msun$ at birth. The lowest mass central BH in a galaxy (ignoring ``normal" stellar mass BHs) has a mass estimate of $\sim$ 50,000 $\msun$ \citep{2015ApJ...809L..14B}, and was identified as a low-luminosity AGN in the dwarf galaxy  RGG~118 \citep{2013ApJ...775..116R,2016ApJ...818..172G,2016MNRAS.460.3119S}. Our implementation allows for such low-mass BHs to be accounted for. 

This implementation is not based on halo properties, but on the local environment of the BH formation sites. The code first identifies all the local overdensities in the gas density field, selects clumps that are bound and collapsing along all axes. If the gas metallicity is below a critical value, they are flagged as potential sites for BH formation. We compute the stellar mass formed in these dense regions and the probability of forming a BH, based on the IMF and the total stellar mass in the clump. Once a region is flagged as a site for BH formation, the mass of the BH is computed, directly related to the stellar mass. Therefore each BH in the simulation box has a different mass, assigned on-the-fly. To mimic the formation of these BHs, we use sink particles, which are able to accrete gas from their surroundings, and to merge together.

SN feedback is of paramount importance, as it modulates metal enrichment, as well as the presence and retention of cold gas in low-mass galaxies. This in turn affects the gas supply to BH seed, and their ability to grow. We compare three implementations of SN feedback, in order of increasing strength: one of the simulations uses a thermal feedback, another one a kinetic feedback, while the last one uses a delayed cooling feedback.  Our main results are as follows. 

\begin{itemize}

\item We find that a stronger SN feedback, delayed cooling, produces galaxies with stellar masses closer to those predicted by the relation with halo mass. 

\item We find that with strong SN feedback, more BHs are formed, but their growth is SN-regulated for low-mass galaxies with $M_*<10^9 \, \rm M_\odot$~\citep{2015MNRAS.452.1502D}: SN-driven winds remove dense star-forming gas and stunt BH accretion in galaxies with shallow potential wells.

\item The lower BH masses and lower accretion rates predicted by the simulation with the strongest SN feedback,  delayed cooling, seem to be in better agreement with the paucity of AGN in high redshift galaxies \citep[][and references therein]{2015MNRAS.448.3167W,2015arXiv151200510C}.

\item We provide the probability that a galaxy of a given mass and redshift hosts a BH (Fig.~\ref{fig:galOF}). This information can be used to seed with BHs lower resolution cosmological simulations.

\item The occupation fraction is also used as a diagnostic of BH formation. Our results agree with analytical and semi-analytical studies \citep{volonteri2008,2010MNRAS.408.1139V,Devecchi2012} and with the simulations by \citet{Bellovary2011}, in that all high-mass haloes/galaxies tend to host a BH, but  low-mass haloes/galaxies have a lower probability of hosting a BH. After BH formation stops at $z\sim 6$, at a given galaxy mass the occupation fraction decreases with time because galaxies grow in mass. 

\item We have compared the BH populations from our simulations to a sample of galaxies representative of the local Universe \citep{2015arXiv150806274R} and to LBAs,  local analogs of high redshift LBGs \citep{2011ApJ...731...55J}.  Our simulated BHs connect to the low-redshift observational sample, and span a similar range in accretion properties as LBAs.  
\end{itemize}

 A follow-up paper will be dedicated to more detailed comparison with these observations.

\section{Acknowledgements}
We thank Roberto Gilli for stimulating discussions and the referee for constructive suggestions that improved the clarity of the manuscript.
MV acknowledges funding from the European Research Council under the European 
Community's Seventh Framework Programme (FP7/2007-2013 Grant Agreement no.\ 614199, project ``BLACK'').  
This project has received funding from the European Union's Horizon 2020 research and innovation programme under the Marie Sklodowska-Curie grant agreement No 656428. 
This work was granted access to the HPC resources of TGCC under the allocation x2014046955,  x2015046955 and x2016046955 made by GENCI, and has made use of the Horizon Cluster, hosted by Institut d'Astrophysique de Paris.
We thank St\'ephane Rouberoul for smoothly running the Horizon Cluster.

\bibliographystyle{mn2e}
\bibliography{biblio_complete,biblio_new}

\label{lastpage}
\end{document}